\documentclass[10pt,twocolumn,floatfix,prl,superscriptaddress]{revtex4-2}
\usepackage{amsmath,amssymb,amsthm,mathrsfs,amsfonts,dsfont,amstext} 
\usepackage{textcomp,pbox}
\usepackage[export]{adjustbox}
\usepackage{bm}
\usepackage{dcolumn,booktabs,url}
\usepackage[scaled]{helvet}
\usepackage{sansmath,gensymb}
\usepackage{tikz,graphicx,transparent,color}
\usepackage{multirow}
\usepackage[separate-uncertainty = true]{siunitx}
\usepackage{comment}
\usepackage{physics}
\usepackage{pdfpages}
\usepackage{array}

\makeatletter
\AtBeginDocument{\let\LS@rot\@undefined}
\makeatother

\newcolumntype{C}[1]{>{\centering\let\newline\\\arraybackslash\hspace{0pt}}m{#1}}

\usepackage[colorlinks=true]{hyperref}
\usepackage{graphicx}

\hypersetup{
     colorlinks   = true,
     citecolor    = blue,
     linkcolor    = blue,
     urlcolor     = blue     
}

\graphicspath{{./figs/}}

\newcommand{\figref}[1]{\figurename{~\ref{#1}}}

\newcommand{\VHOM}{95.7\pm0.8}

\newcommand{\Fcorr}{67.8\%}

\newcommand{\down}{\downarrow}
\newcommand{\up}{\uparrow}
\newcommand{\Down}{\Downarrow}
\newcommand{\Up}{\Uparrow}

\newcommand{\ketUp}{{\ket{\Uparrow}}}
\newcommand{\ketDown}{{\ket{\Downarrow}}}

\newcommand{\fidelity}{\mathcal{F}}

\begin{document}
\newcommand{\TitleName}{Entangling a Hole Spin with a Time-Bin Photon: A Waveguide Approach for Quantum Dot Sources of Multi-Photon Entanglement}

\title{\TitleName}

\newcommand{\AffCPH}{Center for Hybrid Quantum Networks (Hy-Q), The Niels Bohr Institute, University~of~Copenhagen,  DK-2100  Copenhagen~{\O}, Denmark}
\newcommand{\AffBasel}{Department of Physics, University of Basel, Klingelbergstra\ss e 82, CH-4056 Basel, Switzerland}
\newcommand{\AffBochum}{Lehrstuhl f\"ur Angewandte Fest\"orperphysik, Ruhr-Universit\"at Bochum, Universit\"atsstra\ss e 150, 44801 Bochum, Germany}

\author{Martin Hayhurst Appel}
\affiliation{\AffCPH}
\author{Alexey Tiranov}
\affiliation{\AffCPH}
\author{Simon Pabst}
\affiliation{\AffCPH}
\author{Ming Lai Chan}
\affiliation{\AffCPH}
\author{Christian Starup}
\affiliation{\AffCPH}

\author{Ying Wang}
\affiliation{\AffCPH}
\author{Leonardo Midolo}
\affiliation{\AffCPH}
\author{Konstantin Tiurev}
\affiliation{\AffCPH}

\author{Sven Scholz}
\affiliation{\AffBochum}
\author{Andreas D. Wieck}
\affiliation{\AffBochum}
\author{Arne Ludwig}
\affiliation{\AffBochum}
\author{Anders Søndberg Sørensen}
\affiliation{\AffCPH}
\author{Peter Lodahl}
\affiliation{\AffCPH}

\email[Email to: ]{martin.appel@nbi.ku.dk, lodahl@nbi.ku.dk}

\date{\today}

\begin{abstract}
Deterministic sources of multi-photon entanglement are highly attractive for quantum information processing but are challenging to realize experimentally. In this paper, we demonstrate a route towards a  scaleable source of time-bin encoded Greenberger-Horne-Zeilinger and linear cluster states from a solid-state quantum dot embedded in a nanophotonic crystal waveguide. By utilizing a self-stabilizing double-pass interferometer, we measure a spin-photon Bell state with $(67.8\pm0.4)\%$ fidelity and devise steps for significant further improvements. By employing strict resonant excitation, we demonstrate a photon indistinguishability of  $(\VHOM)\%$, which is conducive to fusion of multiple cluster states for scaling up the technology and producing more general graph states.

\end{abstract}

\maketitle 
A near-deterministic source of multi-photon entanglement will enable significant advancements in quantum information processing such as one-way quantum computing consuming few-photon Greenberger-Horne-Zeilinger (GHZ) states \cite{Gimeno-Segovia2015,Rudolph2017} or one-way photonic repeaters relying on large photonic graph states \cite{Borregaard2019, Azuma2015}.
Until now, spontaneous parametric down-conversion has been the leading source of entangled photons and has enabled demonstrations of 12 entangled photons \cite{Zhong2018}. However, this approach is based on probabilistic photon sources and is thus challenging to scale up.
An alternative approach proposed by Lindner and Rudolph exploits a single solid-state quantum emitter as an on-demand entanglement source \cite{Lindner2009}.
In their proposal, a semiconductor quantum dot (QD) containing a single spin is repeatedly excited to create a linear cluster state of polarization encoded photons. Despite its elegance, this approach has several experimental drawbacks. Specifically, natural spin precession is employed in a weak external magnetic field for which the QD spin is poorly decoupled from the nuclear spin bath \cite{Stockill2016,Prechtel2016,Huthmacher2018}. Additionally, the need to collect two orthogonally polarized modes is often experimentally incompatible with the resonant excitation required for high photon indistinguishability~\cite{He2013}.  These challenges were partly combated in the landmark experiment of Ref. \cite{Schwartz2016}, where the use of dark-exciton states and non-resonant excitation enabled the explicit demonstration of three-qubit entanglement and extrapolation to five qubits. Nonetheless, the further scalability of multi-photon entanglement sources has been an open question so far only addressed theoretically \cite{Tiurev2021, Tiurev2022_Short}. A key feature is the generation of high-fidelity multi-photon entanglement while preserving photon indistinguishability; a crucial requirement for fusion-based photonic quantum computing \cite{Gimeno-Segovia2015,Uppu2021a}.

In this paper, we perform the first demonstration of QD spin-photon entanglement using a time-bin protocol capable of generating multi-photon GHZ and linear cluster states. This protocol has been used to demonstrate spin-photon entanglement from nitrogen-vacancy centers \cite{hensen_loophole-free_2015,Tchebotareva2019,Vasconcelos2020}, the first steps using a micropillar QD source were implemented in Ref. \cite{Lee2019} without proving entanglement, and a detailed theoretical treatment was recently published \cite{Tiurev2021,Tiurev2022_Short}. 
Contrary to the Lindner--Rudolph protocol, we apply strong magnetic fields which permit the use of spin-echo pulses and render the protocol insensitive to spin-dephasing~\cite{Tiurev2021}. Additionally, a single optical transition is excited resonantly and emits highly indistinguishable photons into a single-mode photonic crystal waveguide (PCW) with near-unity internal collection efficiency \cite{Arcari2014, Lodahl2015}. Here, we focus on measuring spin-photon Bell states using a novel double-pass interferometer and determine a fidelity of \Fcorr. This fidelity is predominantly limited by the quality of the spin rotations, which can be substantially improved in future experiments~\cite{Bodey2019a}.
\begin{figure*}[t!]%
\centering
\includegraphics[width=1\textwidth]{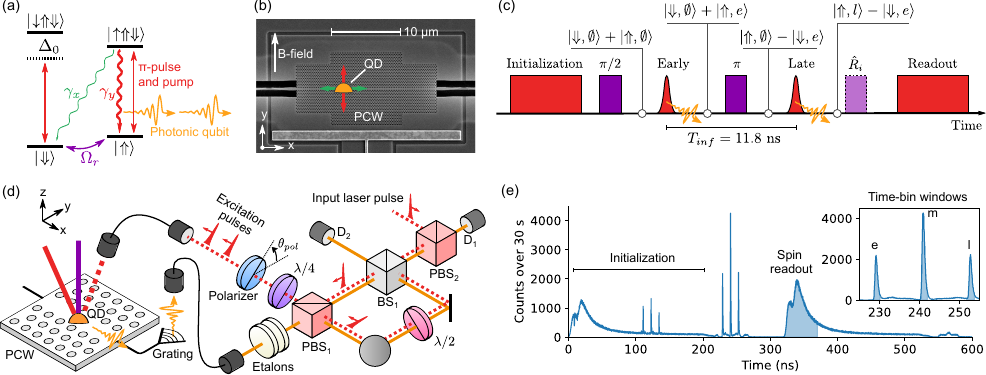}
    \caption{(color online) Time-bin entanglement protocol. 
    \textbf{(a)} Energy level diagram of a positively charged QD in a Voigt magnetic field. The $\ketUp\leftrightarrow\ket{\uparrow\Uparrow\Downarrow}$ transition is used to emit photonic qubits and to perform spin initialization and readout via optical pumping. The trion decay paths have rates $\gamma_y$ and $\gamma_x$, $\Omega_r$ is the spin Rabi frequency and $\Delta_0$ is the sum of the Zeeman splittings. 
    \textbf{(b)} Scanning electron micrograph of the PCW with QD dipoles overlaid. 
    \textbf{(c)} Experimental pulse sequence consisting of pumping pulses (red squares), rotation pulses (purple squares) and fast optical $\pi$-pulses (red pulses). The spin-photon state is indicated on top with $\ket{\emptyset}$ denoting photon vacuum and $\ket{e}(\ket{l})$ denoting an early(late) photon. 
 \textbf{(d)} Experimental setup. The PCW embedded QD is subjected to lasers propagating from free space. A double-pass TBI defines the excitation pulses (dashed red line) and interferes the emitted photons (solid orange line) resulting in mutual phase stability. A polariser with angle $\theta_{pol}$ determines the photonic readout basis. $\lambda/2(\lambda/4)$ denotes half(quarter)-waveplates.
\textbf{(e)} Fluorescence histogram resulting from Bell state generation summed over both detectors including a 50 ns spin readout detection window (shaded area). The inset shows a magnified view of the early(e), middle(m), and late(l) detection windows (2 ns each) which herald the measurement basis of the photonic qubit. The peaks at 120 ns are optical reflections inside the TBI.
}
    \label{fig:1}
\end{figure*}
\\ \\
The entanglement protocol builds upon our recent work on uniting optical cycling transitions with optical spin control~\cite{Appel2021} as summarised in \figref{fig:1}a. A positively charged InAs QD in an in-plane (Voigt geometry) magnetic field $B_y=\SI{2}{T}$ gives rise to a pair of Zeeman-split heavy-hole ground states $\ketUp$ and $\ketDown$, which are used as a spin qubit~\cite{Warburton2013}, and a pair of Zeeman-split trions $\ket{\up\Up\Down}$ and $\ket{\down\Up\Down}$ enabling photon generation. The four linear dipoles are driven by a red-detuned Raman laser to rotate the spin qubit \cite{Bodey2019a, Appel2021} while the PCW (\figref{fig:1}b) selectively enhances the optical transitions resulting in the optical cyclicity $C=\gamma_y/\gamma_x=14.7\pm0.2$~\cite{Appel2021}, which is otherwise unity in a homogeneous environment or a cylindrically symmetric cavity. Spin-photon entanglement is then generated according to the protocol in \figref{fig:1}c. Optical pumping initializes $\ketDown$ and a $\hat{R}_y(\pi/2)$ rotation around the y-axis prepares the superposition state $(\ketDown+\ketUp)/\sqrt{2}$. The QD is then subjected to a y-polarised optical $\pi$-pulse resonant with $\ketUp\leftrightarrow\ket{\uparrow\Uparrow\Downarrow}$. As the other y-polarised transition is detuned by $\Delta_0=2\pi\times\SI{17}{GHz}$, trion excitation and thus photon emission is conditioned on $\ketUp$. The enhanced cyclicity ensures photon emission via the spin-preserving $\ket{\up\Up\Down}\rightarrow\ketUp$ transition with probability $C/(C+1)\approx94\%.$ Thus, an early excitation entangles $\ketUp$ with an early photon $\ket{e}$. The spin states are then swapped by a $\pi$-rotation before applying a late excitation pulse resulting in the emission of a late photon $\ket{l}$. The resulting state is the spin-photon Bell state 
\begin{align}
\ket{\psi_{Bell}}=(e^{i\phi_e}\ket{\Up,l}-\ket{\Down,e})/\sqrt{2},\label{eq:BellState}
\end{align}
where the emission time of the single photon is maximally entangled with the hole spin, and $\phi_{e}$ is the phase difference between the two excitation pulses. 
By working in the rotating frame of the Raman laser, the spin does not precess and only rotates when we actively apply a Raman pulse in contrast to Refs.~\cite{Schwartz2016,Lindner2009}.

The entangled state is generated and analyzed using the setup in \figref{fig:1}d. The QD is held at 4 K in a closed cycle cryostat and is subjected to lasers propagating from free space. The early and late excitation pulses are generated by injecting a single pulse into the excitation-pass (dashed line in \figref{fig:1}d) of a time-bin interferometer (TBI) which sets the phase $\phi_e$ and the time delay $T_{inf}=\SI{11.8}{ns}$ between the early and late pulses. $T_{inf}$ is sufficient for performing photon emission ($400$ ps lifetime) and a $\pi$-rotations (7 ns). The QD emits photons into the guided PCW mode, which is coupled to a single-mode fiber via a grating coupler \cite{Zhou2018}. From here the photon stream enters the detection-pass of the TBI (orange line in  \figref{fig:1}d) where a pair of 3 GHz FWHM etalon filters are used to reject the Raman laser scatter and the QD phonon sideband \cite{Kirsanske2017}. Next, a polarising beamsplitter PBS$_1$ transmits or reflects with equal probability. This passive routing leads to three detection windows (see \figref{fig:1}e) which herald the photonic measurement basis. An early(late) detection corresponds to an early(late) photon propagating through the short(long) TBI arm and thus constitutes a Z-basis measurement. By contrast, the middle window represents the time-bin photon interfering with itself on BS$_1$. A photon click on detector D$_1$ or D$_2$ thus projects the photon onto $\ket{\phi^\pm}=(\ket{e}\pm e^{i\phi_d}\ket{l})/\sqrt{2}$ where $\phi_d$ is the phase of the detection-pass. Since excitation and detection-passes use the same interferometer the phases $\phi_d$ and $\phi_{e}$ are mutually stable ~\cite{Jayakumar2014} giving a stable detection pattern without the need for active stabilization \cite{Vedovato2018, Tchebotareva2019}.

\begin{figure}[t]%
\centering
\includegraphics[width=1\columnwidth]{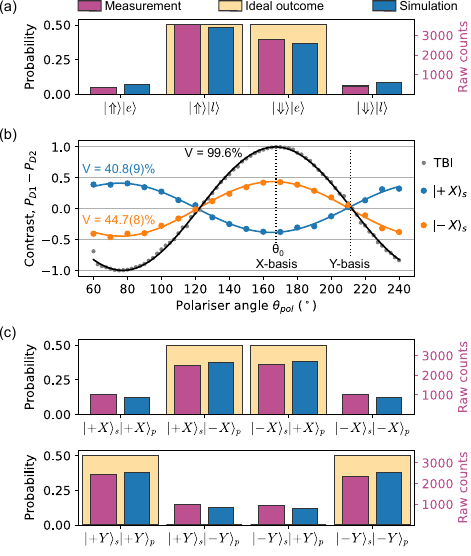}
    \caption{(color online) Entanglement verification. \textbf{(a)} Spin-photon Bell state measured in ZZ-basis. Magenta bars denote the measured spin-photon coincidences during a 120 s acquisition. The right y-axis indicates raw counts and the left y-axis shows the probability normalized across all four outcomes. The yellow and blue bars represent the ideal and simulated detection patterns, respectively.
   \textbf{(b)} Determination of the photonic readout basis. The y-axis indicates the normalized intensity difference between the two TBI detectors in the middle window. The gray dots show the TBI response to a classical input while the blue and orange curves denote spin-photon coincidences when projecting the spin on $\ket{\pm X}_s$ (legend). The fits follow $V\cos(2(\theta_{pol}-\theta_0))$. Error bars are within the data points. \textbf{(c)}  Spin-photon measurements in the XX-basis (top row) and YY-basis (bottom row). The s and p subscripts denote spin and photonic qubit, respectively. Axes and legends follow (a).}
    \label{fig:2}
\end{figure}

We now proceed to quantify the spin-photon entanglement using the approach of Ref. \cite{Guhne2007} which is scalable to future, large $N$-qubit GHZ states by only requiring $N+1$ measurement settings for an exact fidelity estimate. Adapting the technique to a Bell state gives
\begin{align}
\mathcal{F}_{Bell}=\frac{\expval*{\hat{P}_z}}{2}+\frac{\expval*{\hat{M}_y}-\expval*{\hat{M}_x}}{4},\label{eq:Fidelity}
\end{align} 
where $\hat{P}_z=\ketbra{\Up,l}{\Up,l}+\ketbra{\Down,e}{\Down,e}$ measures the classical correlations and  $\hat{M}_x=\hat{\sigma}_x^{(s)}\otimes\hat{\sigma}_x^{(p)}$ and $\hat{M}_y=\hat{\sigma}_y^{(s)}\otimes\hat{\sigma}_y^{(p)}$ measure both qubits along X and Y, respectively. $\hat{\sigma}$ denotes single qubit Pauli matricies, s(p) superscripts denote spin(photonic) qubit, and we designate the logical qubits $\ket{0}=\ketUp,\ket{l}$ and $\ket{1}=\ket\Down,\ket{e}$. 
We post-select on measuring photons in both the photonic and spin readout windows and achieve a 124 Hz coincidence rate when repeating the experiment at a 1.65 MHz repetition rate. As the spin readout can only detect $\ketUp$ we apply a $\hat{R}_i$ rotation prior to readout (\figref{fig:1}c) fulfilling $\hat{R}_i\ket{s}=\ketUp$ to realize the desired spin projector $\ketbra{s}{s}$. \figref{fig:2}a shows the results of a ZZ-basis measurement.
Projection on $\ket{e}$ and $\ket{l}$ is given by the photon detection time (both detectors are treated equally) and the $\hat{R}_i$ pulse is toggled between a 0 and a $\pi$ rotation to realize projections onto $\ketUp$ and $\ketDown$, respectively. Normalizing across all four projections yields $\langle\hat{P}_z\rangle=(89.3\pm0.4)\%$.
The imbalance of $\ket\Up\ket{l}$ and $\ketDown\ket{e}$ is primarily a consequence of the imperfect $\hat{R}_i$ spin rotation (discussed later) which reduces the probability of measuring $\ketDown$. 

In order to measure time-bin encoded photons in the equatorial plane of the Bloch sphere, we add a controllable phase difference between the early and late excitation pulses:
The pulses are combined on PBS$_1$ (\figref{fig:1}d), converted to opposite circular polarizations with a $\lambda/4$ plate, and projected onto the transmission axis of a rotatable polarizer, which adds $2\theta_{pol}$ to the phase $\phi_e$. 
The effect of $\theta_{pol}$ is evident from the gray data in \figref{fig:2}b where classical excitation pulses are reinjected into the detection-pass. This reveals a full oscillation in the detector contrast with near-perfect visibility. The angle $\theta_{pol}=\theta_0$ gives bunching on detector D$_1$ and corresponds to the condition $\phi_d=\phi_e$. $\theta_0$ depends on the specific TBI alignment but is stable on a week-long timescale. We then run the entanglement protocol, project the QD spin onto $\ket{\pm X}_s=(\ketUp\pm\ketDown)/\sqrt{2}$ (for $\hat{R}_i=\hat{R}_y(\pm\pi/2)$) and measure the spin state dependent contrast, see \figref{fig:2}b. Crucially, the two fringes are perfectly in-phase and out-of-phase with the classical TBI response. In summary, measuring the photon in the X-basis corresponds to setting $\theta_{pol}=\theta_0$ and assigning the photon states $\ket{+X}_p$ and $\ket{-X}_p$ to a middle window detection on D$_1$ and D$_2$, respectively. \figref{fig:2}c shows the outcomes of an XX-basis measurement. The outcomes are normalized as before and the contrast between the positive and negative eigenstates of $\hat{M}_x$ results in $\expval*{\hat{M}_x}=(-42.3\pm1.1)\%$. The $\hat{M}_y$ measurement is similarly realized by setting $\theta_{pol}=\theta_0+\pi/4$ and $\hat{R}_i=\hat{R}_x(\pm\pi/2)$ leading to $\expval*{\hat{M}_y}=(42.1\pm1.1)\%$. 
By applying Eq. (\ref{eq:Fidelity}) we arrive at the final estimate $\fidelity_{Bell}^{raw} = (65.7\pm0.4)\%$ which exceeds the classical threshold of $50 \%$ by 39 standard deviations using only 6 minutes of acquisition and no corrections. We note that some of the recorded spin-photon coincidences are due to uncorrelated laser leakage. However, this contribution is minor and correcting for the background (as in Ref.~\cite{Vasconcelos2020}) leads to a corrected fidelity of $\fidelity_{Bell}^{corr}=(67.8\pm0.4)\%$. NV centers \cite{Vasconcelos2020, Tchebotareva2019} have produced similar quality Bell states but with greater reliance on background subtraction. In the interest of understanding the relevant error mechanisms we perform a Monte Carlo simulation of the experiment including all errors and backgrounds (Supplementary Material \cite{SM}). This yields the fidelity $\fidelity_{Bell}^{sim}=67.8\%$ (to be compared against $\fidelity^{raw}_{Bell}$) and the detection pattern in \figref{fig:2} which is in good agreement with the experimental values and supports our error analysis.
\\ \\
We now highlight two of the errors limiting $\mathcal{F}_{Bell}$. The dominant error is the quality of Raman pulses used for X and Y rotations of the hole spin. By measuring the dampening of Rabi oscillations between $\ketUp$ and $\ketDown$ we extract a $\pi$-pulse fidelity of $F_\pi=88.5\%$ which is predominantly limited by incoherent spin-flips between the two spin states (Supplementary Material \cite{SM}). Naturally, the time-bin protocol relies on highly coherent spin control, and $F_\pi=88.5\%$ alone limits the Bell state fidelity to $77\%$ according to Monte Carlo simulations. Thus, we may attribute $\sim70\%$ of the measured infidelity to this mechanism. Fortunately, a recent scheme using electron spins and nuclear spin cooling \cite{Bodey2019a} demonstrated $F_\pi=98.8\%$ and could readily be implemented in the experimental protocol. This would increase the achievable fidelity to $\fidelity_{Bell}=97.3\%$ (neglecting other errors). 

\begin{figure}[t]%
\centering
\includegraphics[width=1\columnwidth]{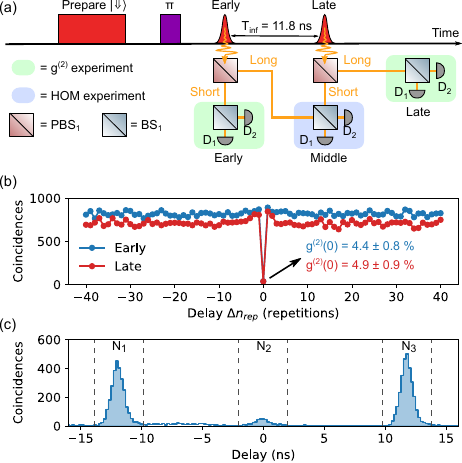}
    \caption{(color online) Single-photon charecterisation.
\textbf{(a)} Pulse sequence for $g^{(2)}$ and HOM measurements.  A pump and a Raman pulse prepare $\ketUp$ and optical $\pi$-pulses are used to emit two photons. The TBI (here in a path-representation) can either measure a single photon on a BS ($g^{(2)}$ measurement) or interfere two photons on a BS (HOM measurement). Waveplates (not shown) ensure identical polarization at the middle detection. \textbf{(b)} $g^{(2)}$ measurement results. The x-axis indicates the delay between photons in units of the experimental repetition time.  \textbf{(c)} Photon coincidences within the same experimental repetition yielding a HOM measurement.}
    \label{fig:3}
\end{figure}

A second relevant error is the single-photon purity and indistinguishability of the emitted photons. The latter error reduces the measurement contrast in the XX and YY bases \cite{Tiurev2021} and is additionally relevant for combining multiple smaller cluster states through entanglement fusion \cite{Rudolph2017}. To accurately characterize the time-bin encoded photon, we retain the magnetic field and minimally modify the pulse sequence to emit two separable single photons (\figref{fig:3}a). By using the TBI, we simultaneously measure the $g^{(2)}$ intensity autocorrelation and Hong--Ou--Mandel (HOM) visibility by letting the detection time herald the experiment. 
\figref{fig:3}b shows the delays between photons recorded in either the early or late detection windows and constitutes two sets of $g^{(2)}$ measurements. A slight bunching is observed for short delays owing to non-deterministic initialization of the hole charge state. Normalizing $g^{(2)}$ at long delays and averaging over early and late gives $g^{(2)}(0)=(4.7\pm0.6)\%$ from which 1.1\% may be attributed to excitation laser scatter. The remaining contribution  is likely a result of multi-photon emission owing to the fast, Purcell enhanced decay rate $\gamma_0=(\gamma_x+\gamma_y)=\SI{2.54}{ns^{-1}}$ and the FWHM duration of the $\pi$-pulse $T_{opt}=\SI{35}{ps}$. $T_{opt}$ represents a trade-off \cite{Tiurev2021,Tiurev2022_Short} between multi-photon emission  (minimized for $T_{opt}\ll\gamma_0^{-1}$) and unwanted excitation of $\ketDown\leftrightarrow \ket{\down\Up\Down}$ (minimized for $T_{opt}\gg\Delta_0^{-1}$). A larger magnetic field will increase $\Delta_0$ and permit a shorter $T_{opt}$ and thus reduced $g^{(2)}(0)$. \figref{fig:3}c shows the delay between two photons recorded within the same experimental repetition when at least one photon was measured in the middle window. Following Ref. \cite{Santori2003}, this estimates a raw indistinguishability of $\mathcal{V}^{raw}=1-2N_2/(N_1+N_3)=(86.5\pm0.6)\%$ (integration windows given in \figref{fig:3}c). $\mathcal{V}^{raw}$ is primarily limited by the finite $g^{(2)}(0)$ according to~ $\mathcal{V}^{raw}\approx\mathcal{V}/(1+2g^{(2)}(0))$, which assumes the multi-photon contribution to consist of distinguishable photons~\cite{Santori2003,Ollivier2021}. Correcting for $g^{(2)}(0)$ and the slight imperfection of the TBI yields a corrected HOM visibility of $\mathcal{V}=(\VHOM)\%$ which is compatible with the QD state of the art \cite{Ding2016, Tomm2021, Uppu2020}.
\\ \\
In summary, we have used a PCW embedded QD to implement a scalable protocol for the generation of time-bin entangled photonic states. This is facilitated by the PCW platform which offers a compelling marriage of spin control and photonic enhancement. Operating at high magnetic fields allows spin initialization without projective measurements and the photon indistinguishability is independent of the magnetic field strength as a single optical transition is used for emission. Our insights from theory and simulation show a clear path towards improving the fidelity with the quality of spin rotations requiring the most attention. Indeed, given realistic PCW parameters and perfect Raman pulses we expect to reach an error level of 2.1\% per emitted photon \cite{Tiurev2022_Short}. The generalization to more photons is straightforward and only requires additional rotation and excitation pulses to create a multi-photon GHZ or 1D-cluster state with photonic qubits emitted every 28 ns. We have attempted a three-qubit GHZ state (Supplementary Material \cite{SM}) but only measured a $\fidelity_{GHZ}=(42.3\pm1.4)\%$ fidelity due to the imperfect Raman pulses. 

Another promising aspect of our approach is the entanglement generation rate. Our 124 Hz Bell-state detection rate is already favourable against similar protocols based on NV centers (7 mHz in Ref. \cite{Vasconcelos2020}) despite our limited and non-optimized total detection efficiency $\eta_{total}=0.3\%$. Indeed, a QD-to-fiber collection efficiency of $\eta=7\%$ was recently demonstrated in a similar PCW structure and further realistic improvements may facilitate collection efficiencies as high as $\eta=78\%$ \cite{Uppu2020}. 
Finally, we note that the high magnetic field regime can give access to nuclear magnon modes \cite{Jackson2021}, which may be used as a long-lived quantum memory for repeater applications \cite{sharman_quantum_2021} or as an ancillary qubit for use in photonic graph state generation \cite{Buterakos2017,Russo2019}.

\begin{acknowledgments}
The supporting data for this Letter are openly available from \cite{Erda}.
\\
The authors thank Alisa Javadi, Matthias C. Löbl and Richard J. Warburton for valuable discussions.  
We gratefully acknowledge financial support from Danmarks Grundforskningsfond (DNRF 139, Hy-Q Center for Hybrid Quantum Networks), Styrelsen for Forskning og Innovation (FI) (5072-00016B QUANTECH), the European Union’s Horizon 2020 research and innovation programme under grant agreement No. 820445 (project name Quantum Internet Alliance), the European Union's Horizon 2020 Research and Innovation programme under grant agreement No. 861097 (project name QUDOT-TECH). 
S. S., A. D. W., and A. L. gratefully acknowledge financial support from Deutsche Forschungsgemeinschaft (DFG) (TRR 160 and LU2051/1-1), the BMBF through project QR.X grant No.  16KISQ009 and the DFH/UFA through grant No. CDFA-05-06.
\end{acknowledgments}

\bibliography{EntanglementPaper_arxiv_final.bbl}
\newpage

\clearpage

\end{document}


\renewcommand*{\Affilfont}{\small}
\date{}
\newcommand{\TitleName}{Supplementary material for <<Entangling a Hole Spin with a Time-Bin Photon: A Waveguide Approach for Quantum Dot Sources of Multi-Photon Entanglement>>}
\title{\TitleName}


\newcommand{\AffCPH}{Center for Hybrid Quantum Networks (Hy-Q), The Niels Bohr Institute, University~of~Copenhagen,  DK-2100  Copenhagen~{\O}, Denmark}
\newcommand{\AffBasel}{Department of Physics, University of Basel, Klingelbergstra\ss e 82, CH-4056 Basel, Switzerland}
\newcommand{\AffBochum}{Lehrstuhl f\"ur Angewandte Fest\"orperphysik, Ruhr-Universit\"at Bochum, Universit\"atsstra\ss e 150, 44801 Bochum, Germany}

\author[1]{Martin Hayhurst Appel}
\affil[1]{\AffCPH{}}
\author[1]{Alexey Tiranov}
\author[1]{Simon Pabst}
\author[1]{Ming Lai Chan}
\author[1]{Christian Starup}
\author[1]{Ying Wang}
\author[1]{Leonardo Midolo}
\author[1]{Konstantin Tiurev}
\author[2]{Sven Scholz}
\affil[2]{\AffBochum{}}
\author[2]{Andreas D. Wieck}
\author[2]{Arne Ludwig}
\author[1]{Anders Søndberg Sørensen}
\author[1]{Peter Lodahl}

\maketitle


\tableofcontents

\newpage
\section{Sample}
\label{sec:sample}
\begin{figure}[h!]
\begin{center}
\includegraphics[width=1\textwidth]{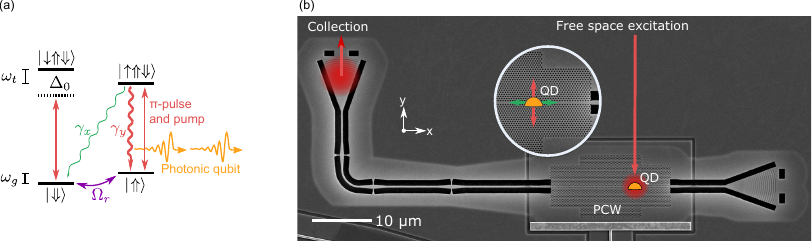}
\caption{\textbf{(a)} Energy level diagram of a positively charged QD in a Voigt magnetic field. $\Down,\Up$ denote hole spins and $\up,\down$ denote electron spins. $\omega_{g(t)}$ is the ground(trion) Zeeman splitting and $\Delta_0=\omega_g+\omega_t$. $\gamma_x,\gamma_y$ are radiative decay rates and $\Omega_r$ is the effective spin Rabi frequency given by the Raman laser. \textbf{(b)} SEM image of the measured two-sided photonic crystal waveguide structure. The QD is addressed by lasers propagating from free space and the emitted photons are coupled into the waveguide mode, scattered out of the left grating coupler and coupled into a single-mode fiber.}
\label{fig:Sample}
\end{center}
\end{figure}

\figref{fig:Sample} shows the quantum dot (QD) level diagram and an SEM picture of the measured waveguide device. The same QD was characterized in Ref. \cite{Appel2021} where additional spectroscopic details are available. Details on sample growth and fabrication are presented in Ref. \cite{Uppu2020}. Key parameters of the positively charged QD are summarized in Table \ref{tab:QDParameters}.

\begin{table}[h]
\centering
\begin{tabular}{|l|l|}
\hline
\textbf{Property}                      & \textbf{Value}                      \\ \hline
Emission wavelength           & 945.0 nm                   \\ \hline
Trion decay rate $\gamma_0=\gamma_x+\gamma_y$  & $(2.48\pm0.02)$ ns$^{-1}$ \\ \hline
Optical cyclicity $C=\gamma_y/\gamma_x$             & $14.7\pm0.2$               \\ \hline
Transform-limited linewidth $\gamma/(2\pi)$   & $(395\pm3)$ MHz                    \\ \hline
Measured linewidth            & 1050 MHz                   \\ \hline 
Trion Zeeman splitting $\omega_t/(2\pi)$       & 9.6 GHz                  \\ \hline
Electron g-factor             & 0.34                       \\ \hline
Ground state Zeeman splitting $\omega_g/(2\pi)$ & 7.3 GHz                 \\ \hline
Hole g-factor                 & 0.26                       \\ \hline
Spin dephasing time $T_2^*$           & $(23.2\pm1.4)$ ns                    \\ \hline
\end{tabular}
\caption{Key properties of the studied positively charged QD. All parameters are measured in a $B_y=\SI{2}{T}$ magnetic field.}
\label{tab:QDParameters}
\end{table}

\newpage
\section{Experimental Setup}
\label{sec:setup}
\begin{figure}[h!]
\begin{center}
\includegraphics[width=1\textwidth]{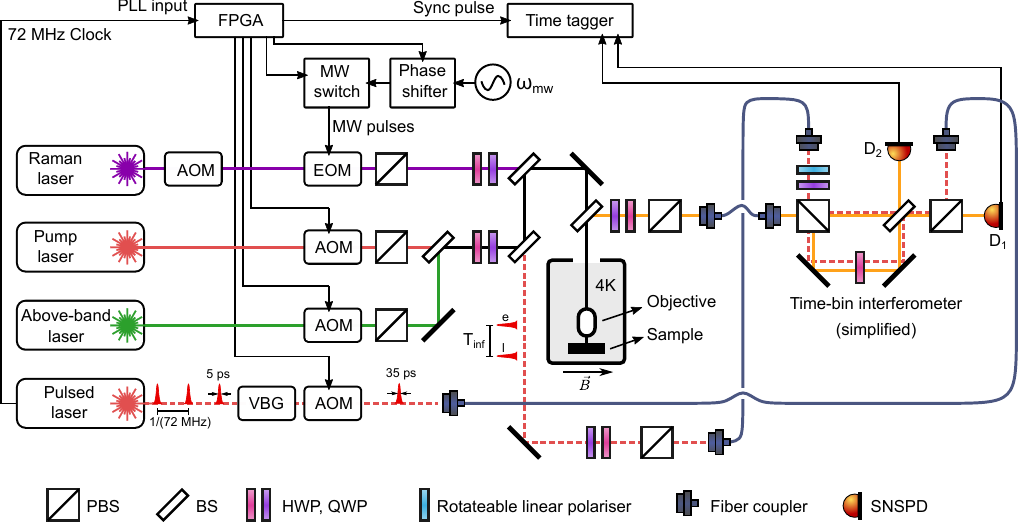}
\caption{ Experimental setup.}
\label{fig:Setup}
\end{center}
\end{figure}

An overview of the experimental setup is given in \figref{fig:Setup}. Using a series of beamsplitters (BS), four lasers are combined and reflected into the cryostat. A polarizing beamsplitter (PBS) and a pair of half-waveplates (HWP) and quarter-waveplates (QWP) control the laser polarization in each path. Additionally, each laser path contains an acousto-optical modulator (AOM) to control the optical power. The four lasers comprise
\begin{enumerate}
    \item Raman laser used for spin rotations. The CW laser is amplitude modulated by a fiber-coupled electro-optical modulator (EOM) to add sidebands matching the hole spin Zeeman splitting \cite{Bodey2019a}. By combining a microwave source, a variable phase shifter and a microwave of/off-switch, we create microwave pulses that drive the EOM. Additionally, an AOM is used to stabilize the optical power.
    \item Pumping laser used for spin initialization and readout. The CW laser is modulated by an AOM to create pulses (8 ns rise time).
    \item 830 nm laser used to initialize the hole charge state. The CW laser is modulated by an AOM to create pulses (30 ns rise time). A 100 ns pulse is placed at the end of the entanglement pulse sequence to initialize a hole spin for the next round of the experiment.
    \item Pulsed laser used for optical $\pi$-pulses. A mode-locked Ti-Sapphire laser creates 5 ps pulses at a 72 MHz repetition rate. The pulses are stretched in time to $T_{opt}=35$ ps using a volume Bragg grating (VBG) and an AOM is used to pick out a single pulse. This pulse is doubled using the excitation-pass of the TBI (detailed in \sectionref{sec:tbi}) before being combined with the other lasers.
\end{enumerate}
As the pulsed laser has a fixed repetition rate it constitutes the experimental clock to which all other devices are phase-locked. This is achieved with a custom FPGA solution. A photodiode detects the 72 MHz pulse train and drives a phase-locked loop (PLL) in the FPGA. The PLL drives a series of pulse generators within the FPGA which deliver TTL pulses to the modulators and the time tagger (see \figref{fig:Setup}).

Photon detection is achieved with a pair of superconducting nanowire single-photon detectors (SNSPDs) with 70-90\% quantum efficiency, and a Swabian Ultra time tagger is used to record the photon detection times and the FPGA synchronization pulses. Detector dark counts $\sim1$ Hz are completely negligible. 
Detector time jitter ($\sim$ 100 ps RMS) is irrelevant as the time-bin protocol only requires the photonic detection windows (2 ns duration) to be resolved.

The sample itself is held at 4.2 K inside a closed-cycle cryostat. A superconducting vector magnet provides the 2 T in-plane magnetic field. The sample is imaged with a 0.81 NA objective and 3 piezo positioners are used to translate the sample and bring it into focus of the objective. Additionally, a DC voltage source provides a bias voltage \bias to the sample. This in conjunction with the 830 nm laser facilitates positive charging of the QD \cite{Appel2021}.

\clearpage
\section{Time-Bin Interferometer}
\label{sec:tbi}
\begin{figure}[h!]
\begin{center}
\includegraphics[width=0.8\textwidth]{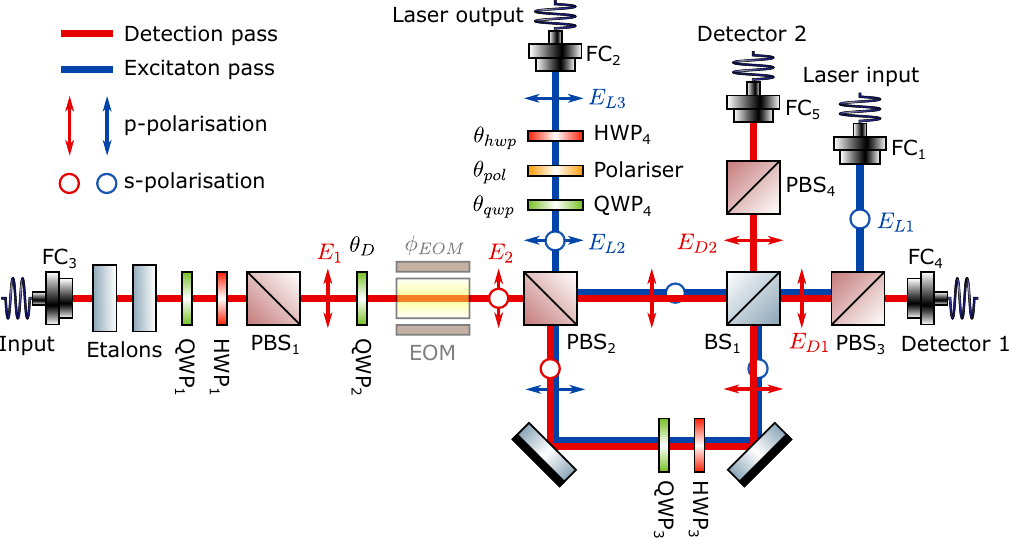}
\caption{Double pass time-bin interferometer. The excitation-pass (blue) doubles the laser pulse inserted in FC$_1$. 
The detection-pass (red) applies frequency filtering and interferes early and late pulses leading to detection in FC$_4$ or FC$_5$. QWP$_2$ and a free space EOM set the splitting ratio of PBS$_2$ and may be used for active photon routing. Lenses and additional mirrors are excluded for clarity.} 
\label{fig:TBI}
\end{center}
\end{figure}

The full layout of the time-bin interferometer (TBI) is shown in \figref{fig:TBI}. It contains a detection-pass for measuring the time-bin photon and an excitation-pass for generating the early and late excitation pulses. The long arm of the interferometer comprises a 3.54 m free space delay resulting in a $T_{inf}=\SI{11.8}{ns}$ delay.

\subsection{Working Principle}
The detection-pass begins with two identical etalons with 3 GHz linewidth and 100 GHz free spectral range. The etalons are resonant with the $\ketUp\leftrightarrow\ketUpe$ transition and transmit the time-bin photon but provide rejection of the Raman laser (spectrally located between two longitudinal etalon modes), phonon sideband, and emission from the other three optical transitions of the QD. After the etalons, a pair of waveplates and PBS$_1$ fix the photon polarization. Next, a free-space EOM and QWP$_2$ set the splitting ratio of PBS$_2$. The EOM can be used to enable active switching but was not driven in the reported measurement. Instead, the EOM provided a constant birefringence and QWP$_2$ was rotated to achieve a 50/50 splitting ratio of PBS$_2$. After PBS$_2$ the time-bin photon interferes with itself on BS$_1$. Waveplates in the long arm ensure identical polarization of the beams impeding on BS$_1$. Finally, the time-bin photon is coupled into fiber couplers FC$_4$ and FC$_5$ which lead to two SNSPD detectors.

In the excitation-pass a single laser pulse is emitted from FC$_1$ and injected into the TBI via PBS$_3$. On encountering BS$_1$ the pulse is divided. The long interferometer arm creates the late pulse which becomes p-polarized thanks to HWP$_3$ and QWP$_3$. After being recombined by PBS$_2$, the early and late pulses have electric field amplitudes proportional to
\begin{align}
\mathbf{E}_{L2}^e&=\sqrt{T_\text{short}}\jonesVec{0}{1},\label{eq:ExcitaionEfield1}\\
\mathbf{E}_{L2}^l&=\sqrt{T_\text{long}}e^{i\phi^\prime}\jonesVec{1}{0},\label{eq:ExcitaionEfield2}
\end{align}
where the $e(l)$ superscripts denote early(late), and $T_\text{short(long)}$ is the intensity transmission coefficients of the short(long) arm including the BS$_1$ splitting ratio and other losses. $\phi^\prime$ is the relative phase accumulated in the long arm and $\smqty(1\\0)$ and $\smqty(0\\1)$ denotes p- and s-polarization, respectively. The orthogonal polarizations enable us to set a programmable phase shift using QWP$_4$, a linear polarizer and HWP$_4$. 
These three elements are described by the total Jones matrix~\cite{Hecht2002}
\begin{align}
M_{tot} &= M_{hwp}(\theta_{pol}/2)M_{pol}(\theta_{pol})M_{qwp}(\theta_{qwp})\\
&=\left(\frac{1+i}{2}\right)\begin{bmatrix}
\cos(\theta_{pol})-i\cos(\theta_{pol}-2\theta_{qwp})&\sin(\theta_{pol})+i\sin(\theta_{pol}-2\theta_{qwp})\\
0&0
\end{bmatrix},\label{eq:PolariserTransform}
\end{align}
where $\theta_{pol}$ and $\theta_{qwp}$ denote the polarizer transmission axis and QWP fast axis with respect to horizontal, respectively. The HWP angle is set to always ensure a p-polarized output. 
Multiplying \eqref{eq:PolariserTransform} onto \eqref{eq:ExcitaionEfield1} and \eqref{eq:ExcitaionEfield2} gives the electric fields at FC$_2$
\begin{align}
E_{L3}^e&=\left(\frac{1+i}{2}\right)\sqrt{\Tshort}(\sin(\theta_{pol})+i\sin(\theta_{pol}-2\theta_{qwp}))\label{eq:LaserEout1},\\
E_{L3}^l&=\left(\frac{1+i}{2}\right)e^{i\phi^\prime}\sqrt{\Tlong}(\cos(\theta_{pol})-i\cos(\theta_{pol}-2\theta_{qwp})).
\label{eq:LaserEout2}
\end{align}
By choosing $\theta_{qwp}=\pi/4$ and $T_{short}=T_{long}$, the two pulses achieve equal intensity and a phase difference of 
\begin{align}
\phi_e=\arg(E^e_{L3})-\arg(E^l_{L3})=\phi^\prime +2\theta_{pol}-\pi/2.
\end{align}
In essence, the detection pattern depends on the phase of the detection interferometer $\phi_d$ and the phase of the excitation interferometer $\phi_e$. They are mutually phase-stable, as the two passes encounter the same optics. However, by scanning $\theta_{pol}$ an additional controllable phase can be added, which is enough to effectively scan the interferometer.

In practice, we achieve $T_{short}=T_{long}$ by manipulating the optical coupling into FC$_2$ and exploiting that the short and long paths do not have perfect spatial overlap at FC$_2$.

The excitation and detection-passes are in general orthogonally polarized and displaced 3 mm vertically resulting in low levels of cross-talk. We estimate that only one in $\approx2\cdot 10^{11}$ excitation photons are reflected into a detector. This is sufficient, although some reflections are visible in the entanglement data (main text Fig. 1e ).

\subsection{Interferometer Characterization}
\begin{figure}[h]
\begin{center}
\includegraphics[width=1\textwidth]{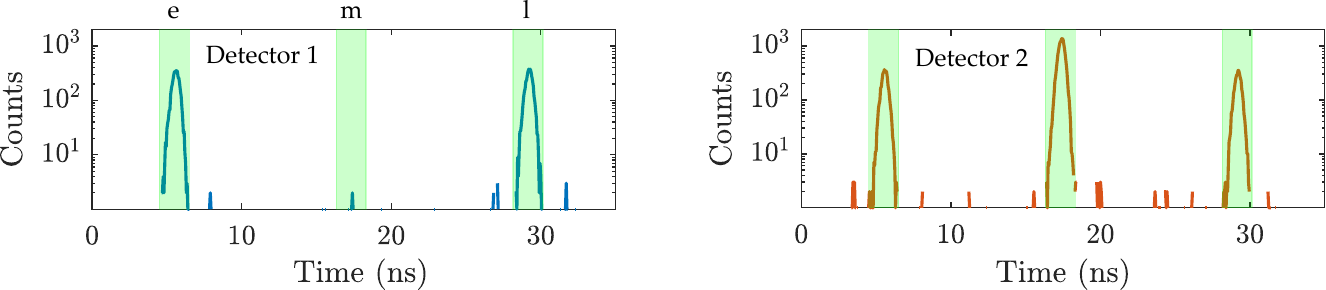}
\caption{Histograms recorded from both TBI detectors showing excitation pulses reflected of the sample. The early(e), middle(m) and late(l) detection windows are 2 ns long and marked in green. The applied $\theta_{pol}$ setting corresponds to series 1 minimum in \figref{fig:DetectionInterferometerVsPolariser} and yields a near-zero intensity in the middle window of detector 1.}
\label{fig:Interferometer:ThetaDScan}
\end{center}
\end{figure}

\begin{figure}[h!]
\begin{center}
\includegraphics[width=0.9\textwidth]{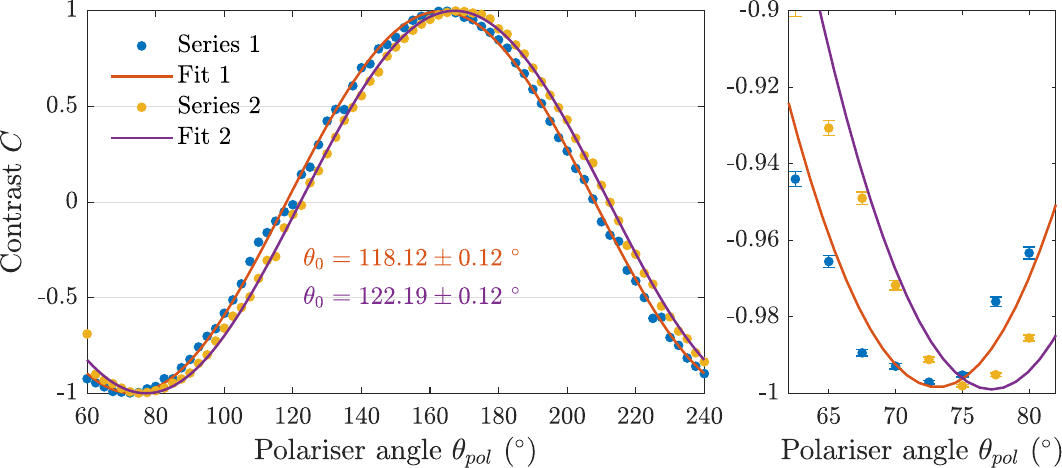}
\caption{Detector contrast for different values of $\theta_{pol}$. The right subplot shows a data zoom in. Errorbars are derived from shot noise. Fits following \eqref{eq:TBIFit} are used to estimate the fringe phase and the TBI visibility.}
\label{fig:DetectionInterferometerVsPolariser}
\end{center}
\end{figure}

We now verify the stability and visibility of the TBI. To do so, a pair of early/late excitation pulses is generated by the TBI, reflected of the sample surface and coupled into the TBI detection-pass. This procedure is equivalent to connecting fiber couplers FC$_2$ and FC$_3$ in \figref{fig:TBI}. We define the detector contrast $C=(I_{1}^m-I_{2}^m)/(I_{1}^m+I_{2}^m)$ where $I_{i}^m$ is the intensity of detector $i$ within the middle detection window. \figref{fig:Interferometer:ThetaDScan} shows how the intensities are extracted from a histogram and  \Figref{fig:DetectionInterferometerVsPolariser} shows $C(\theta_{pol})$ for two separate measurements taken 142 hours apart without intermediate alignment. The detector contrast is fit with the model
\begin{align}
C(\theta_{pol})=\nu_{TBI}\sin(2(\theta_{pol}-\theta_0))\label{eq:TBIFit}
\end{align}
and estimates near-perfect visibilities of $\nu_{TBI}=99.7\%$ and $\nu_{TBI}=99.9\%$ for series 1 and series 2, respectively. 
Furthermore, the fringe phase $\theta_0$ only shifts by $(4.1\pm 0.2)^\circ$ between the two measurements. The $5$ minutes required to calibrate $\theta_0$ are negligible compared to the duration of stability, and the TBI is thus self-stabilizing for our purposes. Slight systematic deviations from the fit are visible in \figref{fig:DetectionInterferometerVsPolariser} owing to a $\sim5\%$ variation in the relative intensity of the early and late pulses as a function of $\theta_{pol}$. 

%
%
%
%
%

%
%

\clearpage
\section{Optical Loss budget}
Table \ref{tab:LossBudget} gives an estimated loss budget for the entanglement experiment and predicts a total loss of 20.2 dB. In comparison, the $g^{(2)}$ experiment (main text) yields a 0.282\% (25.5 dB loss) single-photon detection efficiency, and we thus have 5 dB of unaccounted loss. These losses may be due to device fabrication imperfections such as over-etched grating couplers \cite{Zhou2018}. The two-sided nature of the PCW and the 50/50 BS in the collection path (\figref{fig:Setup}) constitute the main losses. Additional improvements may be gained by adding distributed Bragg reflectors to the gratings as discussed in Ref. \cite{Uppu2020}.
\begin{table}[h]
\centering
\begin{tabular}{llll}
\textbf{Loss source}            & \textbf{\begin{tabular}[c]{@{}l@{}}Efficiency \\ (\%)\end{tabular}} & \textbf{\begin{tabular}[c]{@{}l@{}}Loss \\ (dB)\end{tabular}} & \textbf{Source/assumptions}                                     \\ \hline
\hline
\textbf{Sample}                         &                          &                    &                                                             \\
Hole spin initialization                 & 75                       & 1.25               & Blinking measurements  \\
Zero-phonon line emission               & 90                       & 0.46               & \cite{Lodahl2015} \\
Branching efficiency\tabfoot{1}  & 88                       & 0.56               & \cite{Appel2021}      \\
Two-sided PCW                   & 50                       & 3.01               & Equal left/right coupling                   \\
Nanobeam interfaces             & 96                       & 0.18               & \cite{Uppu2020}                            \\
Grating coupler diffraction efficiency\tabfoot{2}               & 60                       & 2.22               & \cite{Uppu2020}                           \\\hline 
Sum                             &                          & \textbf{7.68}               &                                                             \\ \hline
                                &                          &                    &                                                             \\
\textbf{Cryostat optics}                 &                          &                    &                                                             \\
Optical inset\tabfoot{3}                  & 75                       & 1.25               & Measured                                                    \\
50/50 BS                       & 49                       & 3.10               & Measured                                                    \\
Polarization optics & 85                       & 0.71               & Measured                                                    \\
Collection fiber coupling        & 60                       & 2.22               & Measured      \\ \hline
Sum                             &                          & \textbf{7.27              } &                                                             \\ \hline
                                &                          &                    &                                                             \\
\textbf{Interferometer}   &                          &                    &                                                             \\
Two fiber matings               & 76                       & 1.19               & Measured                                                    \\
Etalon filtering\tabfoot{4}                & 73                       & 1.37               & Measured                                                    \\
Interferometer optics           & 80                       & 0.97               & Measured                                                    \\
Interferometer fiber coupling         & 84                       & 0.76               & Measured                                                    \\
SNSPD detectors                 & 80                       & 0.97               & Manufacturer specifications                                \\ \hline
Sum                             &                          & \textbf{5.25              } &                                                             \\ \hline
                                &                          &                    &                                                             \\ \hline
Total loss                      &                          & \textbf{20.2             } &                                                             \\ \hline
\end{tabular}
\caption{Estimated experimental loss budget.
\tabfoot{1} The probability of the photon being emitted via the cycling transition and being emitted into the waveguide mode.
\tabfoot{2} The probability of the photon being scattered into the microscope objective. This does not include the mode overlap with the collection fiber. 
\tabfoot{3} Microscope objective and lenses in 4F-system.
\tabfoot{4} Includes scattering losses and spectral overlap between a zero-phonon line photon and the etalon system. Phonon sideband losses are noted separately. 
}
\label{tab:LossBudget}
\end{table}

\clearpage
\section{Spin Dependent Optical Excitation}
\label{sec:excitation}
We now explicitly demonstrate that the detection of a photon is strongly conditioned on the QD spin state. The same magnetic field $B_y=\SI{2}{T}$ is applied as in the main experiment and the spin is prepared in $\ketDown(\ketUp)$ by optically pumping into $\ketDown$ and applying a $0(1\pi)$ spin rotation. The QD is then excited by a single 35 ps excitation pulse resonant with $\ketUp\leftrightarrow\ketUpe$ and the QD emission intensity is measured using the short arm of the TBI. 

\begin{figure}[h!]
\begin{center}
\includegraphics[width=0.7\textwidth]{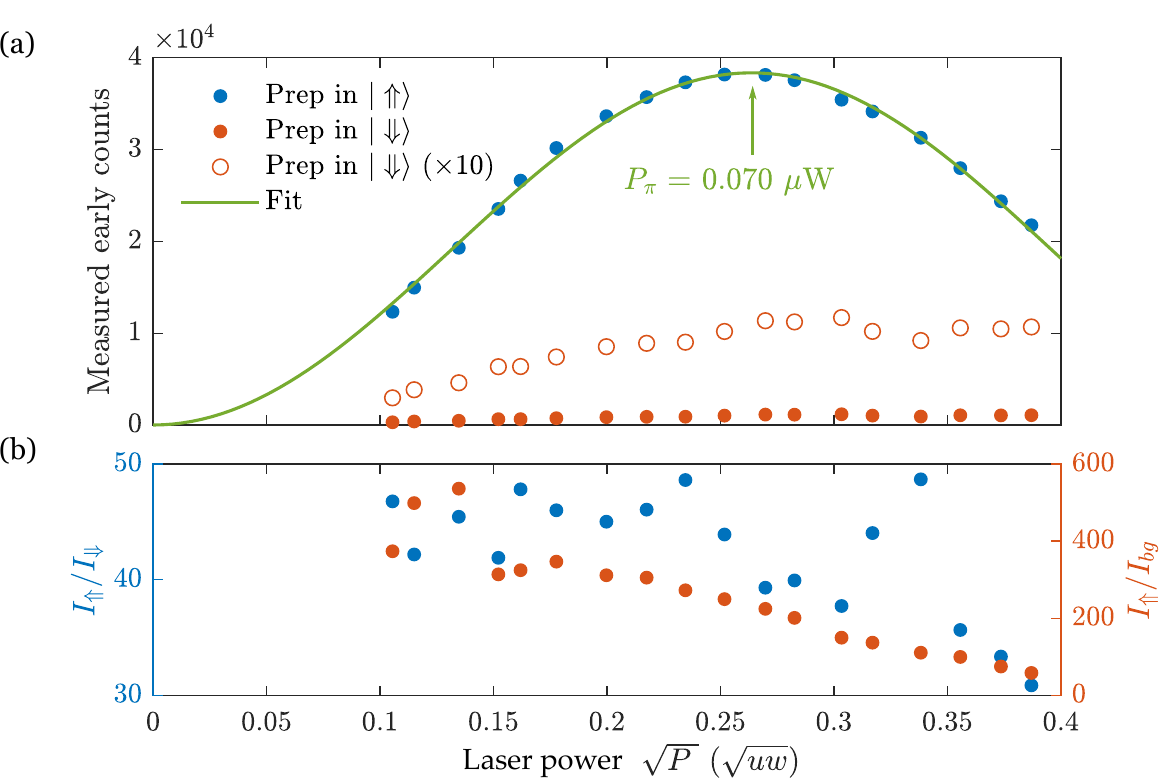}
\caption{QD fluorescence under pulsed excitation as a function of laser pulse power $P$ and QD spin state using a $B_y=\SI{2}{T}$ magnetic field. \textbf{(a)} Spin state dependent fluorescence. The fit follows \eqref{eq:RabiFit}. \textbf{(b)} Left y-axis: Ratio between fluorescence intensities conditioned on spin state. Right y-axis: Fluorescence given $\ketUp$ divided by laser background. }
\label{fig:OpticalExci}
\end{center}
\end{figure}

\figref{fig:OpticalExci}a shows the emission intensity as a function of spin state and excitation power $P$. Given initialization in $\ketUp$, the measured intensity $I_\Up$ is well fit by two-level Rabi-flopping~\cite{Gerry2005}
\begin{align}
I_\Up(P)=I_{max}\times\sin\left(\frac{\pi}{2}\sqrt{\frac{P}{P_\pi}}\right)^2,\label{eq:RabiFit}
\end{align}
where $I_{max}$ is the maximum intensity and $P_\pi$ is the $\pi$-pulse power. The fit in \figref{fig:OpticalExci}a provides an estimate of $P_\pi$ which is used for the remaining experiments

Preparing $\ketDown$ results in the intensity $I_\Down\ll I_\Up$ . For the five lowest powers, the ratio of intensities has an average value $\langle I_{\Uparrow}/I_{\Downarrow} \rangle\approx 45$ which is consistent with the limit set by the spin initialization fidelity. 

Additionally, the laser background $I_{bg}$ can be estimated by applying a non-resonant \bias which renders the QD optically inactive. The laser extinction ratio $I_\Up/I_{bg}$ is given in \figref{fig:OpticalExci}b and reaches a value of $\approx 240$ at $P=P_\pi$. The extinction is sensitive to the exact optical alignment and varies slightly between data sets.

In summary, the QD emission under pulsed excitation is seen to be highly spin-dependent, has the dynamics of a single two-level system, and exhibits low levels of laser background.
\newpage
\section{Spin Control}\label{sec:spincontrol}
\begin{figure}[h!]
\begin{center}
\includegraphics[width=0.5\textwidth]{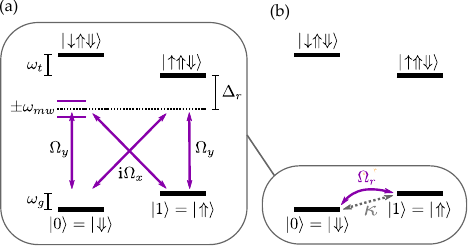}
\caption{Raman level scheme. \textbf{(a)} A detuned, circularly polarized laser drives a two-photon Raman transition between the heavy-hole ground states of a positively charged QD. $\Delta_r$ is the optical detuning, $\omega_g(\omega_t)$ is the ground state(trion) Zeeman splitting and $\Omega_x,\Omega_y$ are the optical Rabi frequencies of the $x$ and $y$-dipoles. \textbf{(b)} Effective spin dynamics with spin Rabi frequency $\Omega_r$ and spin-flip rate $\kappa$. }
\label{fig:spin_control} 
\end{center}
\end{figure}
Spin control follows the level scheme in \figref{fig:spin_control}. A circularly polarized Raman laser drives all four linear optical dipoles and is detuned by $\Delta_r=2\pi\times \SI{350}{GHz}$ from $\ket{\up\Up\Down}$.
An amplitude modulating EOM (\figref{fig:Setup}) driven at frequency $\omega_{mw}$ creates two sidebands detuned $\pm \omega_{mw}$ from the fully suppressed carrier. The coherent evolution of the ground states is given by~\cite{Bodey2019a}
\begin{align}
\hat{H}_{rot}&=\mqty[-\delta/2 & \frac{\Omega_r}{2}e^{-2i\phi_{mw}} \\
\frac{\Omega_r}{2}e^{2i\phi_{mw}} & \delta/2],\label{eq:Heff4}
\end{align}
where $\Omega_r=\Omega_x\Omega_y^*/\Delta_r$ is the real, effective spin Rabi frequency, $\delta=\omega_g-2\omega_{mw}$ is the two-photon detuning and $\phi_{mw}$ is the modulation phase. By toggling a phase switch with the FPGA (\figref{fig:Setup}), $\phi_{mw}$ can be switched between two arbitrary values on a ns-timescale. This provides control over the rotation axis of the final $\hat{R}_i$ rotation pulse. 

\subsection{Quality of Rotation Pulses}\label{sec:RotationFidelity}
\begin{figure}[h!]
\begin{center}
\includegraphics[width=1\textwidth]{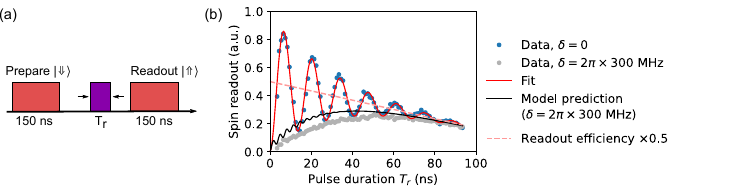}
\caption{Spin Rabi oscillations. \textbf{(a)} Experimental pulse sequence. \textbf{(b)} Intensity of spin readout proportional to the population of $\ketUp$. The pulse duration is varied for two values of $\delta$. The dampened oscillations are fitted (red line) with \eqref{eq:SpinControl:EmpericalFitModel}. The dashed line indicates the modelled readout efficiency.}
\label{fig:SpinRabi} 
\end{center}
\end{figure}
To estimate the rotation pulse fidelity we employ the prepare-rotate-readout sequence of \figref{fig:SpinRabi}a to observe the spin Rabi oscillations in \figref{fig:SpinRabi}b. We employ the same optical power and $\Delta_r$ as in the entanglement experiment. The measurement series with $\delta=0$ exhibits clear dampened Rabi oscillations which we fit with a master equation including effects from both inhomogeneous broadening and spin-flips.
We represent the heavy-hole spin states $\{\ketDown,\ketUp\}$ as $\{\ket{0},\ket{1}\}$ and take the evolution of the density matrix $\hat{\rho}$ to follow the Lindblad master equation
\begin{align}
\frac{d}{dt}\hat{\rho}=-i[\hat{H}_{rot}, \hat{\rho}] +\sum_{j=1}^2 \left[\Chat_j \hat{\rho}\Chat_j^\dag -\frac{1}{2}(\Chat_j^\dag \Chat_j\hat{\rho} + \hat{\rho} \Chat_j^\dag \Chat_j)\right],\label{eq:MasterEquation}
\end{align}
where $\hat{H}_{rot}$ is given by \eqref{eq:Heff4}. Spin-flips are incorporated via the collapse operators $\Chat_1=\sqrt{\kappa}\ketbra{0}{1}$ and $\Chat_2=\sqrt{\kappa}\ketbra{1}{0}$ where $\kappa$ is the spin-flip rate. $\kappa$ is assumed equal for both operators as the Zeeman energy $\hbar\omega_g=h\times\SI{7.3}{GHz}$ is small compared to the thermal energy $k_BT\approx h\times\SI{80}{GHz}$. \eqref{eq:MasterEquation} gives rise to the equations of motion
\begin{align}
\frac{d}{dt}\mqty(\rho_{00}\\\rho_{01}\\\rho_{10}\\\rho_{11}) =
\mqty(
-\kappa & \frac{i\Omega_r}{2} & -\frac{i\Omega_r}{2} & \kappa \\
\frac{i\Omega_r}{2} & -\kappa+i\delta & 0 & -\frac{i\Omega_r}{2}\\
-\frac{i\Omega_r}{2} & 0 & -\kappa -i\delta & \frac{i\Omega_r}{2} \\
\kappa & -\frac{i\Omega_r}{2} & \frac{i\Omega_r}{2} &-\kappa
)
\mqty(\rho_{00}\\\rho_{01}\\\rho_{10}\\\rho_{11}),
\label{eq:SpinControl:SpinMasterequation}
\end{align}
for $\phi_{mw}=0$. Using the initial state $\hat{\rho}=\ketbra{0}{0}$, \eqref{eq:SpinControl:SpinMasterequation} is integrated over the pulse duration $T_r$ to yield the $\rho_{11}(T_r)$ population. We assume the sidebands to be on resonance such that $\delta\rightarrow \delta_{OH}$ where $\delta_{OH}$ is the two-photon detuning caused by the quasi-static \cite{Urbaszek_nuclear_2013} nuclear Overhauser field\footnote{In addition to nuclear noise, charge noise also gives a slow inhomogeneous broadening of hole spins \cite{Huthmacher2018,Prechtel2016}. However, the measurement performed here can not distinguish between the two.}. This broadening is incorporated by performing the integral
\begin{align}
\expec{\rho_{11}(T_r;\Omega_r,\kappa)}_{inhom}=\int\displaylimits_{-\infty}^\infty d\delta_{OH}
\rho_{11}(T_r;\Omega_r,\kappa,\delta_{OH})\frac{e^{-\delta_{OH}^2/(2\sigma^2)}}{\sigma\sqrt{2\pi}},\label{eq:SpinControl:rho11Numerical}
\end{align}
where $\sigma=\sqrt{2}/T_2^*$ is the RMS fluctuation of $\delta_{OH}$ and is fixed according to the independently measured $T_2^*=\SI{23.2\pm1.3}{ns}$ which is largely comparable with hole spins in bulk QDs \cite{Godden2012,Delteil2016, Huthmacher2018}. Finally, the master equation is related to the measured intensity via 
\begin{align}
I(T_r)=I_0(1-\alpha T_r)\expec{\rho_{11}(T_r;\Omega_r,\kappa)}_{inhom},\label{eq:SpinControl:EmpericalFitModel}
\end{align}
where $\alpha$ is an empirical parameter associated with a $T_r$-dependent reduction in readout efficiency. The model provides an excellent fit of the data (red line in \figref{fig:SpinRabi}b) and estimates $\Omega_r=2\pi\times\SI{72}{MHz}$ and $\kappa=\SI{0.018}{ns^{-1}}$. Evaluating $\rho_{11}$ after a $1\pi$-oscillation estimates a $\pi$-pulse fidelity of $F_\pi=88.5\%$. The infidelity owing to inhomogeneous broadening can be quantified according to $1-F_\pi^{deph}\approx 1/(\Omega_rT_2^*)^2=1.8\%$. Hence, we attribute the remaining $\approx10\%$ infidelity to $\kappa$, which is by far the main cause of dampening. In summary, we distinguish between $\kappa$ and $\delta_{OH}$ induced dampening by applying a master equation approach, fixing $T_2^*$ based on independent measurements and keeping $\kappa$ as a free fitting parameter.

We justify the inclusion of $\alpha$ by noting that spin readout for $\delta=2\pi\times\SI{300}{MHz}$ (\figref{fig:SpinRabi}b) starts to decrease for long $T_r$. For $\delta\gg\Omega_r$, the dynamics are dominated by spin-flips and populations evolve monotonously according to \eqref{eq:SpinControl:SpinMasterequation}. Thus, the observed decrease in $I(T_r)$ cannot be attributed to spin-flips and is thus an indication of a modulated readout efficiency. This is compatible with the "power tuning" observed in all our devices \cite{Ding2019} where an increase in optical power shifts the optical resonances, possibly due to optically generated charge carriers. The choice of a linearly decreasing readout efficiency is somewhat ad-hoc and leads to negative efficiencies for long $T_r$ but has the benefit of only adding one additional fitting parameter. Using the fitted values of $\Omega_r,\kappa,\alpha$ and setting $\delta=2\pi\times\SI{300}{MHz}$ produces the black curve in \figref{fig:SpinRabi}b, which shows limited agreement with data for short $T_r$. This motives further measurements of $\kappa$ but does not alter the conclusion, namely that $T_2^*$ alone cannot explain the poor $\pi$-pulse fidelity which considerably impacts the entanglement experiment (see \sectionref{sec:montecarlo}).

%
%
%
\section{Hong-Ou-Mandel Visibility}
\label{sec:hom}
We now elaborate on the estimation of the Hong-Ou-Mandel (HOM) visibility. Here, the spin is prepared in $\ketUp$ and the QD is excited twice resulting in two separable photons (early and late) within the 606 ns repetition time. $g^{(2)}$ and HOM are estimated from the same 20-minute acquisition.

Most HOM estimates rely on varying the degree of two-photon interference, either by varying the temporal overlap \cite{Hong1987} or by varying the photon polarization \cite{Uppu2020}. Since the TBI does not give access to these degrees of freedom, we instead follow the method of Ref.~\cite{Santori2003} in which two photons are injected into an unbalanced Michelson interferometer that closely resembles the TBI.
The raw HOM visibility is calculated from
\begin{align}
\vhom^{raw}=1-\frac{N_{m1,m2}}{(N_{m1,e2}+N_{m2,e1}+N_{m1,l2}+N_{m2,l1})/2},\label{eq:HOMRaw}
\end{align}
where $N_{ij,kl}$ is the number of coincidences between a photon in detection window $i$ and detector $j$ and a second photon in detection window $k$ and detector $l$, with both photons being detected within the same experimental repetition. In other words: Coincidences between both detectors at the middle bin are normalized to the coincidences between a middle and non-middle detection. 

Following Ref. \cite{Santori2003}, we consider two mechanism which may reduce the HOM visibility of a perfect source of indistinguishable photons. Firstly, the signal photons may be mixed with noise photons which are fully distinguishable from the signal photons. Secondly, the detection interferometer may be imperfect. In the presence of these errors an ideal source will result in the measured visibility
\begin{align}
\vhom^{\text{ideal}}=1-\frac{R_P T_P \left[(g^{(2)}+1)(R^2+T^2)-2RT\mathcal{V}_{TBI}^2\right]}{RT(R_P^2+T_P^2)(g^{(2)}+1)},\label{eq:HOMCorrection1}
\end{align}
where we use the shorthand notation $g^{(2)}=g^{(2)}(0)$, $\mathcal{V}_{TBI}$ is the classical visibility of the TBI, R(T) is the reflection(transmission) of BS$_1$ (\figref{fig:TBI}) and $R_P$($T_P$) is the reflection(transmission) of the input photons incident on PBS$_1$ (\figref{fig:TBI}). Reflection and transmission coefficients are with respect to power and are measured with a power meter. \eqref{eq:HOMCorrection1} differs slightly from Ref. \cite{Santori2003} as our interferometer utilizes two beamsplitters with independent splitting ratios. We also emphasize that \eqref{eq:HOMCorrection1} assumes $g^{(2)}>0$ to result from noise photons which are distinguishable from the ideal single photons. This assumption has been experimentally justified in some cases \cite{Ollivier2021} but we do not investigation the assumption here.

The corrected visibility is then given by
\begin{align}
\vhom^{corr}=\frac{\vhom^{raw}}{\vhom^\text{ideal}}.\label{eq:HOMCorrection2}
\end{align}
and represents the HOM visibility in the limit of a perfect interferometer and $g^{(2)}=0$.

The data yield a raw visibility of $\vhom^{raw}=86.5\pm0.7\%$. 
Applying \eqref{eq:HOMCorrection2} and the $\vhom^\text{ideal}$ estimated in Table \ref{Table:HOMCorrections} yields the corrected value $\vhom^{corr}=95.7\pm0.8$. The residual distinguishability can be attributed to elastic phonon scattering~\cite{Tighineanu2018}.
\begin{table}[h!]
\centering
\begin{tabular}{|l|l|l|}
\hline
\textbf{Correction}      & \textbf{Experimental values}                                                                                 & \textbf{\begin{tabular}[c]{@{}l@{}}Visibility reduction\\ $1-\vhom^{ideal}$\end{tabular}}                     \\ \hline
Splitting ratio          & \begin{tabular}[c]{@{}l@{}}$R_P=0.485,\;T_P=0.515$, \\ $R=0.513,\;T=0.487$\end{tabular}             & 0.14\%                                                                                    \\ \hline
Classical TBI visibility & $\mathcal{V}_{TBI}=0.997$                                                                 & 0.60\%                                                                                    \\ \hline
Intensity autocorrelation                 & $g^{(2)}=0.047$ & 9.0\%\\ \hline
Total reduction                    &                                                                                                 & 9.7\%\\ \hline
\end{tabular}
\caption{Correction factors used to calculate $\vhom^{ideal}$.  \label{Table:HOMCorrections}}
\end{table}
\\
Finally, we elucidate a small detail in the $g^{(2)}$ measurement in Fig. 3b of the main text. At long delays the rate of late photon coincidences amounts to 88.4\% of the rate of early photon coincidences. This difference can be attributed to the fact that a late photon is conditioned on the early emission occurring via the cycling transition with probability $C/(C+1)$. The coincidence probability for two late photons is thus scaled by $(C/(C+1))^2=(87.7\pm0.2)\%$ relative to the early case in close agreement with the experiment.

\clearpage
\section{Bell State Analysis}\label{sec:fidelity}
Here we explicitly describe the procedure used to estimate the Bell state fidelity with and without background correction. When analyzing the acquired photon time-tags, we post-select on observing at least one click in the photonic detection windows
and at least one click in the spin readout window. Due to the low detection efficiency, the probability of observing multiple photons in the photonic windows is negligible. 

\subsection{Background Correction}\label{sec:Background}
To estimate a background-free detection pattern, we consider the observed photons to be either signal photons originating from the QD or uncorrelated background photons. We then assume the number of observed coincidences to follow
\begin{align}
N_\text{observed}=N_\text{true}+M_{rep}\big[P(b_p)P(b_r)+P(b_p)P(s_r)+P(s_p)P(b_r)\big],\label{eq:EntanglementBackground}
\end{align}
where $N_\text{true}$ is the number of coincidences owing to two signal photons, $M_{rep}$ is the number of experimental repetitions and the $P()$ terms represent the probability of observing a signal(s) or a background(b) photon occurring in a photonic detection window (p-subscript) or the spin readout window (r-subscript). Thus, the last three terms in \eqref{eq:EntanglementBackground} represent combinations of signal and background resulting in false coincidences. We estimate these terms as $P(a)=N_a/M_{rep}$ where $N_a$ is the number of detections of type $a$. Inserting in \eqref{eq:EntanglementBackground} and rearranging the terms gives an estimate of the true number of coincidences
\begin{align}
\expec{N_\text{true}}=N_\text{observed}-\frac{1}{M_{rep}}\left(N(b_p)N(b_r)+N(b_p)N(s_r)+N(s_p)N(b_r)\right).\label{eq:EntanglementBackgroundSubtraction}
\end{align}

\begin{figure}[h!]
\begin{center}
\includegraphics[width=0.8\textwidth]{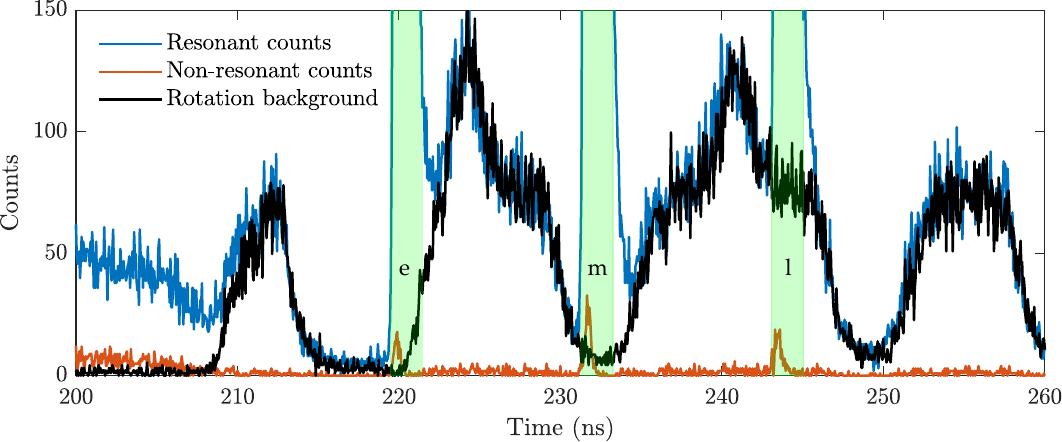}
\caption{Sources of background during the photonic detection windows. Blue (red) curves show fluorescence histograms from the full entanglement experiment using a $\hat{R}_i=\hat{R}_y(\pi)$ spin rotation and a resonant (non-resonant) \bias. The blue curve extends up to  2000(4000) counts during the $e(m)$ window. The black curve shows fluorescence generated by the rotation laser at resonant \bias. All histograms were acquired over 30 s. 
Green windows represent the $e$, $m$, $l$ detection windows. Especially the $l$ window is contaminated by rotation background. } 
\label{fig:RotationLaserBackground}
\end{center}
\end{figure}

We consider three different sources of background photons:
\begin{enumerate}
\item Laser scatter from the excitation laser (already discussed in \sectionref{sec:excitation}). 
\item Raman pulses. The Raman laser is spectrally located at the transmission minimum of the etalons and scatter from this laser is effectively rejected. However, the QD may emit fluorescence during the Raman pulses owing to a finite trion population. This is evident from \figref{fig:RotationLaserBackground} where the Raman pulses are only evident at a resonant \bias. The Raman pulses do not themselves overlap with the photon emission in time. However, when measured through the TBI, the pulses are doubled in time and some overlap occurs. This overlap is most pronounced when projecting the spin onto $\ketDown$ for which a $\hat{R}_i=\hat{R}_y(\pi)$ rotation is applied after the late excitation (see \figref{fig:RotationLaserBackground}). 
\item The pumping laser produces a small amount of laser scatter in the spin readout.
\end{enumerate}

%
%
%
%
%
%
\begin{figure}[h!]
\begin{center}
\includegraphics[width=1\textwidth]{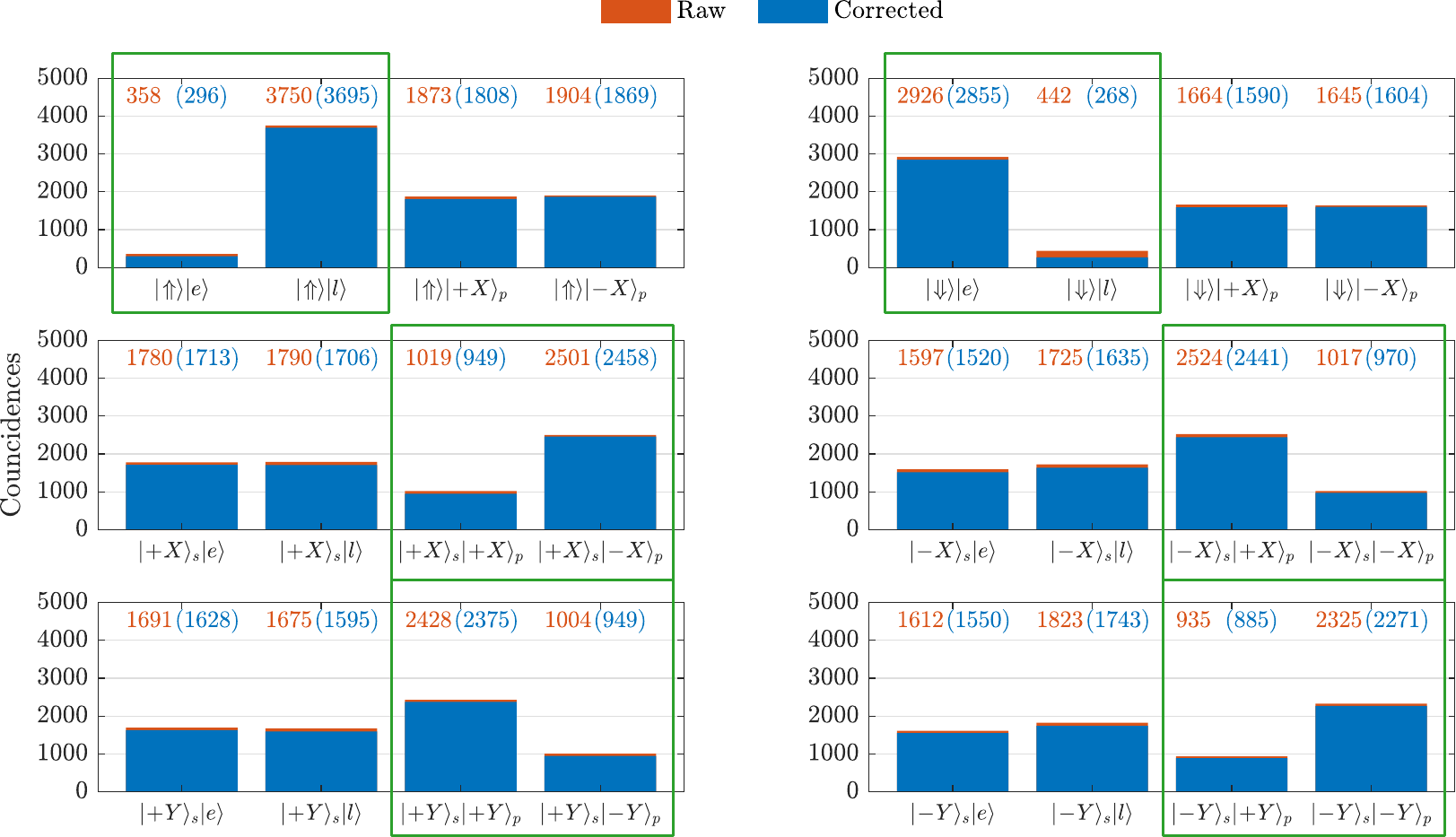}
\caption{Coincidence counts from the six measurement settings used to characterize the Bell state. The x-labels denote the measured spin and photon projectors. 
A 60 s integration time was applied for each of the six measurement setting.} 
\label{fig:GHZHistogram}
\end{center}
\end{figure}

\clearpage
\subsection{Fidelity Estimation}
\figref{fig:GHZHistogram} shows the number of spin-photon coincidences with and without background correction. Measuring the $\hat{P}_z,\;\hat{M}_x$ and $\hat{M}_y$ operators requires six measurement settings as only one spin-projector can be measured at a time. As the photonic basis is selected randomly, only half of the outcomes (green boxes in \figref{fig:GHZHistogram}) are used to construct Fig. 2a,c in the main text. The operators of interest and the Bell state fidelity~\cite{Guhne2007} are calculated according to 

\begin{align}
\expval{\hat{P}_z}&=\frac{N_{\Down,e}+N_{\Up,l}}{N_{\Down,e}+N_{\Up,l}+N_{\Up,e}+N_{\Down,l}},\label{eq:Entanglement:PzDef}\\
\expval{\hat{\mathcal{M}}_x}&=\frac{N_{+X,+X}+N_{-X,-X}-N_{+X,-X}-N_{-X,+X}}{N_{+X,+X}+N_{-X,-X}+N_{+X,-X}+N_{-X,+X}},\label{eq:Entanglement:M1Def}\\
\expval{\hat{\mathcal{M}}_y}&=\frac{N_{+Y,+Y}+N_{-Y,-Y}-N_{+Y,-Y}-N_{-Y,+Y}}{N_{+Y,+Y}+N_{-Y,-Y}+N_{+Y,-Y}+N_{-Y,+Y}},\label{eq:Entanglement:M2Def}\\
\expval{\hat{\chi}}&=\frac{\expval{\hat{\mathcal{M}}_y}-\expval{\hat{\mathcal{M}}_x}}{2},\\
\fidelity_{Bell}&=\frac{\expval{\hat{P}_z}+\expval{\hat{\chi}}}{2},
\end{align}
where the $N$ refers to the number of observations in \figref{fig:GHZHistogram}. The "$+$" and "$-$" photon states correspond to measuring the photon on detector 1 and 2 in the middle detection window, respectively. The photon measurement basis is switched between $X$ and $Y$ by increasing $\theta_{pol}$ by $\pi/4$ corresponding to a $\pi/2$ change of the interferometer phase. The operator estimates and $\fidelity_{Bell}$ are given in Table \ref{tab:fidelity} with and without background subtraction. 

\begin{table}[h!]
\centering
\begin{tabular}{|l|l|l|l|}
\hline
                             & Ideal value & Raw value & Corrected value  \\ \hline
$\expval*{\hat{P}_z}$          	& $1$ 	& $0.893\pm0.004$	&  $0.921\pm0.004$      \\ \hline
$\expval*{\hat{\mathcal{M}}_x}$ &$1$ 	& $0.423\pm0.011$ 	&  $0.437\pm0.011$     \\ \hline
$\expval*{\hat{\mathcal{M}}_y}$ &$-1$ 	& $-0.421\pm0.011$	&  $-0.434\pm0.011$        \\ \hline
$\expval*{\hat{\chi}}$          &$1$	& $0.422\pm0.008$ 	&  $0.436\pm0.008$        \\ \hline
$\fidelity_{Bell}$             	&$100\%$ 	& $(65.7\pm0.4)\%$	&  $(67.8\pm0.4)\%$       \\ \hline
\end{tabular}
\caption{Fidelity estimates with and without background correction. Uncertainties denote $1\sigma$ errors derived from Poissonian statistics.}
\label{tab:fidelity}
\end{table}

\newpage
\subsection{Alternative estimates}
We have thus far proven spin-photon entanglement by estimating the fidelity to a Bell state using a limited set of measurements. This method is motivated by the fact that it will allow efficient fidelity estimates of $N$-qubit GHZ states using only $N+1$ measurement settings (provided both spin-projections can be detected from one setting). However, to facilitate comparison with the existing literature, we also estimate the negativity and concurrence of the measured Bell state. Since the performed measurements do not constitute a full tomography we follow Ref. \cite{Chou2005} and assume the spin-photon density matrix to take the form
\begin{align}
\hat{\rho}=\begin{pmatrix}
\rho_{00,00} & 0 & 0 & \rho_{00,11}e^{i\phi} \\
0 & \rho_{01,01} & 0 & 0 \\
0 & 0 & \rho_{10,10} & 0 \\
\rho_{00,11}e^{-i\phi} & 0 & 0 & \rho_{11,11} 
\end{pmatrix}, \label{eq:rho_model}
\end{align}
where all $\rho$ are real. The diagonal elements are extracted from the top row of \figref{fig:GHZHistogram}. For the off-diagonal elements in \eqref{eq:rho_model}, we use the relation \cite{Guhne2007}
\begin{align}
\expec{\hat{\chi}}=Tr\{\hat{\rho}(\dyad{00}{11}+\dyad{11}{00})\}=2\cos(\phi)\rho_{00,11},\label{eq:rho_bound}
\end{align}
where $\expec{\hat{\chi}}$ is the experimentally measured visibility. Thus, \eqref{eq:rho_bound} provides a lower bound estimate of $\rho_{00,11}\geq \expec{\hat{\chi}}/2$. 

Using the raw, uncorrected measurements we estimate the entanglement concurrence $\mathcal{C}$ and negativity $\mathcal{N}$ according to their standard definitions \cite{Guhne2009} and find
\begin{align}
\mathcal{C}&\geq0.316\pm0.008,\\
\mathcal{N}&\geq0.158\pm0.004,
\end{align} 
with uncertainties calculated from resampling the measurement outcomes. We note that the estimated quantities represent lower bounds, partly due to the lower bound of $\rho_{00,11}$, and partly due to the fact, that additional non-zero off-diagonal elements in \eqref{eq:rho_model} will increase $\mathcal{C}$ and $\mathcal{N}$ without affecting the performed measurements.

\section{Monte Carlo Simulation}
\label{sec:montecarlo}
In this section, we briefly describe the infidelity mechanisms considered in our Monte Carlo (MC) simulation of the Bell state generation. We consider eight errors of which three relate to the intrinsic level structure of the QD. These errors are treated analytically in Refs. \cite{Tiurev2021, Tiurev2022_Short} while Ref. \cite{Appel2021a} discusses the remaining errors in further detail. The errors comprise
\begin{enumerate}
\item Finite cyclicity~\cite{Tiurev2021}. With probability $1/(1+C)$ the $\ket{\up\Up\Down}$ trion wrongly decays "diagonally" to $\ketDown$. The emitted photon is rejected by the etalon filters and is not detected. However, a diagonal decay during the early excitation may result in the emission and detection of a late photon resulting in dephasing of the quantum state.
\item Finite photon indistinguishability ~\cite{Tiurev2021}. The scattering of a phonon during photon emission provides which-path information and dephases the quantum state.
\item Excitation errors \cite{Tiurev2021}. Due to the finite detuning $\Delta_0$ and pulse duration $T_{opt}$, the $\ket{\Down\up\down}$ trion may be excited and emit a photon. The photon is rejected by the etalons but provides which-path information and thus dephasing. Multi-photon emission also results in dephasing.
\item Spin rotation errors. As discussed in \sectionref{sec:RotationFidelity}, $T_2^*$ and spin-flips with rate $\kappa$ result in limited rotation fidelity. The two dampening mechanisms are implemented separately. 
\item Spin initialization. Following the initialization pulse, the spin is the mixture $\hat{\rho}=F_i\ketbra{\Up}{\Up}+(1-F_i)\ketbra{\Down}{\Down}$.
\item Spin readout. Given a photon detection during spin readout, the spin state is $\ketUp$ with probability $F_r$. $F_r$ is limited by laser background and imperfect dynamics, e.g. spin-flips and optical repumping coupling $\ketUp$ and $\ketDown$.
\item Finite classical interferometer visibility $\mathcal{V}_{TBI}$ leads to reduced visibility in XX and YY basis.
\item Uncorrelated counts during the photonic detection windows (discussed in section \ref{sec:Background}).
\end{enumerate}

Additionally, we model the optical loss of the system and use post-selection criteria identical to the experiment. Similar to the experiment, the simulated post-selection does not allow us to distinguish between photons emitted during or after an optical $\pi$-pulse. However, we assume that photons emitted during a $\pi$-pulse suffer an additional 6 dB loss based on their increased bandwidth and the narrowband etalon filters. This slightly increases fidelity, as "bad" trajectories with a photon emitted mid-$\pi$-pulse are less likely to pass post-selection.

Table \ref{table:MonteCarlo:Infidelity} lists the contribution of each error and highlights spin rotations as the dominant error source. Additionally, spin initialization and readout can benefit from improvement but are less critical as their impact remains constant when increasing the number of photons. 

A simulation including all error mechanisms yields $\fidelity^{sim}_{Bell}=67.8\%$ and the detection pattern in Fig. 2 of the main text. This is in good agreement with the measured raw fidelity $\fidelity_{Bell}^{raw} = (65.7\pm0.4)\%$. Note that the infidelity $1-\fidelity^{sim}_{Bell}$ is less than the sum of infidelities in Table \ref{table:MonteCarlo:Infidelity} as some errors will cancel out. 
\begin{table}[h]
\centering\small
\begin{tabular}{lllll}
\hline
Error                                                                            & $\hat{P}_z$                                                  & $\hat{\chi}$ & Parameter value                                                                                 & Infidelity \\ \hline
Cyclicity                                                                        & \begin{tabular}[c]{@{}l@{}}No \\ (for Bell state)\end{tabular}   & Yes          & $C=14.7$ & 1.62\%~(simulated)                      \\ \hline
Phonon dephasing                                                                        & No                                                           & Yes          & $\vhom^{corr}=95.7\%$                                                                                      & 2.34\%~(simulated)                      \\ \hline
Excitation                                                                       & \begin{tabular}[c]{@{}l@{}}No \\ (for $C\gg 1$)\end{tabular} & Yes          & \begin{tabular}[c]{@{}l@{}}$\gamma_0=\gammavalue$,\\ $\Delta_0=17\times 2\pi$,\\$T_{opt}=\Toptvalue$.\end{tabular}          & 2.66\%~(simulated)                       \\ \hline
\begin{tabular}[c]{@{}l@{}}Spin control\\  ($\kappa$ \& $T_2^*$)\end{tabular}    & Yes                                                          & Yes          & \begin{tabular}[c]{@{}l@{}}$T_2^*=\Ttwostarvalue$,\\  $\kappa=\SI{0.021}{ns^{-1}}$\end{tabular} & 22.8\%~(simulated)              \\ \hline
Spin readout                                                                     & Yes                                                          & Yes          & $F_r=96.6\%$  & 5.1\%~(simulated)                        \\ \hline
Spin initialization                                                              & No                                                           & Yes          & $F_i=98\%$    & 2.0\%~(simulated)                          \\ \hline
\begin{tabular}[c]{@{}l@{}}Laser background,\\  uncorrelated counts\end{tabular} & Yes                                                          & Yes          & Background estimates    & 2\%~(calculated)                                                                                                              \\ \hline
Interferometer                                                                   & No                                                          & Yes          & $\mathcal{V}_{TBI}=99.7\%$                                                                      & 0.15\%~(calculated)     \\ \hline
\end{tabular}
\caption{Summary of all considered error mechanism, their key parameter values, and their impact on Bell state fidelity (in absence of other errors) according to MC simulations using $10^6$ trajectories and calculations. The $\hat{P}_z$ and $\hat{\chi}$ columns indicate whether an error reduces visibility of ZZ and XX/YY-basis measurements, respectively. }
\label{table:MonteCarlo:Infidelity}
\end{table}

\clearpage
\section{Three-Qubit GHZ Measurement}\label{sec:3qubit}
\begin{figure}[h!]
\begin{center}
\includegraphics[width=1\textwidth]{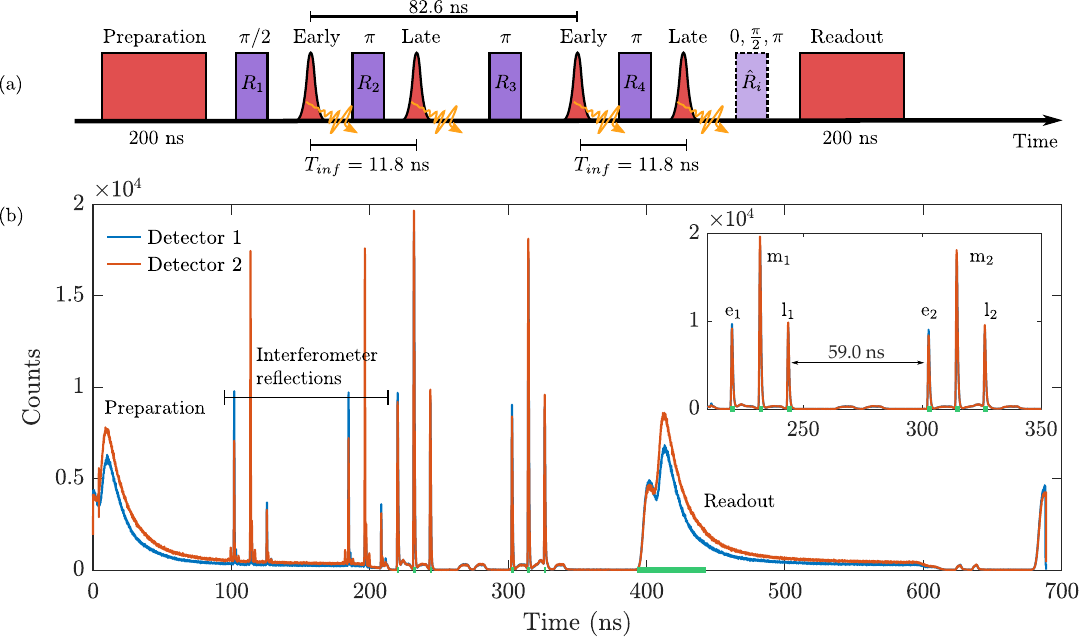}
\caption{Three-qubit experiment. \textbf{(a)} Experimental pulse sequence. Two extra spin rotations ($R_3$ and $R_4$) and a second set of early/late optical excitations are added to generate a second time-bin encoded photon. 
\textbf{(b)} Histogram obtained from 300 seconds of acquisition. The green lines show detection windows. The inset shows a magnified view of the early(e), middle(m) and late(l) detection windows of the two photonic qubits. 
} 
\label{fig:3qubit:Hist}
\end{center}
\end{figure}

As a proof of concept, we expand the experimental pulse sequence to emit a second time-bin entangled photon. Our target state is thus the three-qubit GHZ state
\begin{align}
\ket{\psi_{GHZ}}=\frac{1}{\sqrt{2}}\left(\ket{\Up,l_1,l_2}+\ket{\Down,e_1,e_2}\right),
\end{align}
where $e(l)$ denotes an early(late) photon and the subscripts denote photons in the order of emission. Fig. \ref{fig:3qubit:Hist} shows the full pulse sequence and a recorded histogram. As our setup only includes two detectors we are forced to add an 82.6 ns delay between the generation of the first and second photon to exceed the detector dead time. This extra delay could be removed by using detectors with faster recovery or detector multiplexing. Our post-selection is similar to the two-qubit case: We require at least one detection per photonic qubit and at least one detection from the spin readout. Using this requirement, our coincidence rate drops to 0.68 Hz.
%
%
%
%
%
%
%
\begin{figure}[h!]
\begin{center}
\includegraphics[width=1\textwidth]{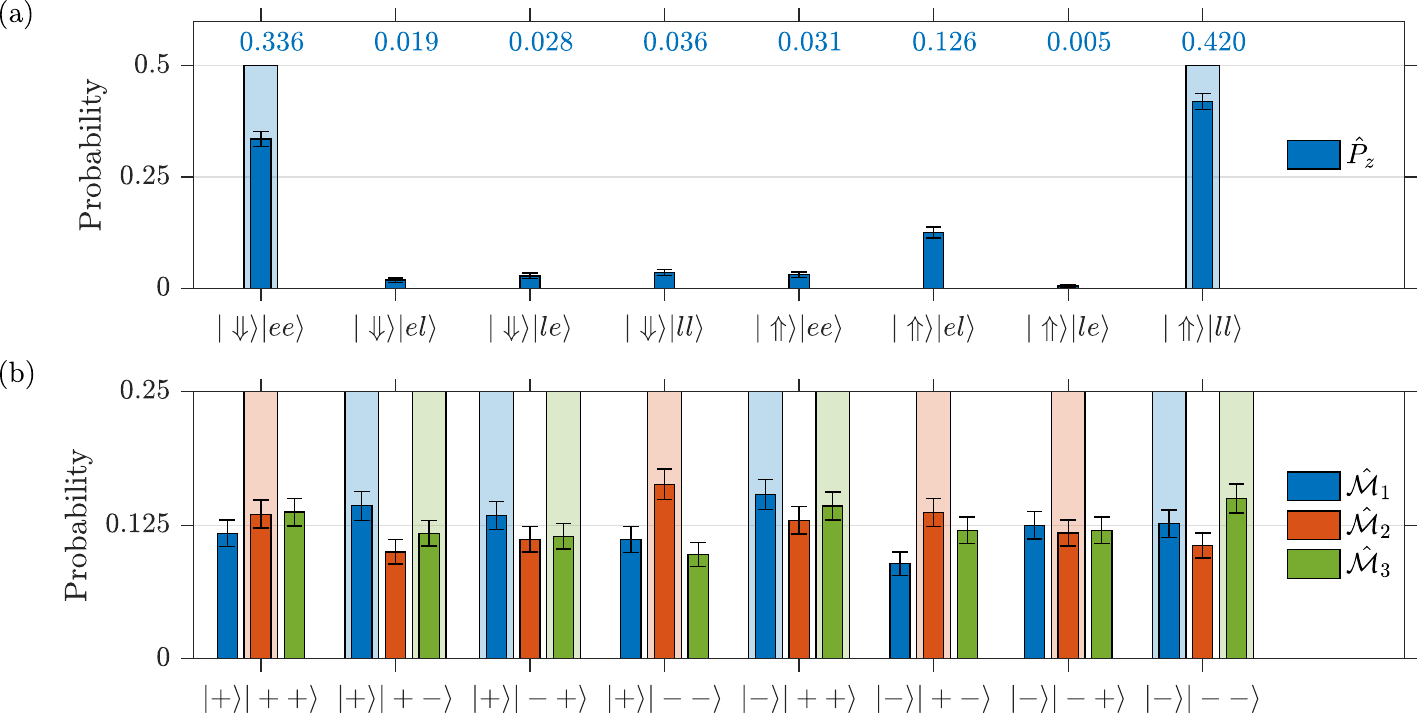}
\caption{Fidelity estimation of a three-qubit GHZ state. Probabilities are normalized across all outcomes. Shaded bars represent ideal outcomes. Errorbars are derived from shot noise.
 \textbf{(a)} Measurements in ZZZ-basis.  \textbf{(b)} Measurements of three equidistant projectors on the Bloch sphere equator. Each state on the x-axis is an eigenstate of $\hat{\mathcal{M}}_i$. The qubits are in the order of spin, photon 1 and photon 2.}
\label{fig:3qubit:GHZ}
\end{center}
\end{figure}

\figref{fig:3qubit:GHZ}a shows the raw histogram for the ZZZ-basis measurement and reveals a detection pattern with $\expval*{\hat{P}_z}=(75.5\pm1.6)$. As expected, this visibility is lower than the two-qubit case as more operations (and especially Raman pulses) are performed.

Estimating $\hat{\chi}$ now requires measuring the operators \cite{Guhne2007}
\begin{align}
\hat{\mathcal{M}}_k &= \left[\cos\left(\frac{k\pi}{N}\right)\hat{\sigma}_x+\sin\left(\frac{k\pi}{N}\right)\hat{\sigma}_y\right]^{\otimes N},\text{  for $k=\{1,2,3\}$}.\label{eq:Mk}
\end{align}
The raw measurement outcomes are given in \figref{fig:3qubit:GHZ}b and give rise to the visibilities in Table \ref{tab:3qubit}. We note that $\expval*{\hat{\chi}}$ is larger than zero by more than three standard deviations and thus represents a significant coherence. However, we do not exceed the $\fidelity=0.5$ threshold for entanglement.

\newcommand{\newrow}{\\ \hline}
\begin{table}[h]
\centering
\renewcommand{\arraystretch}{1.7}
\begin{tabular}{|l|l|l|}
\hline
                               & Ideal value & Estimated value \newrow
$\expval*{\hat{P}_z}$           & 1 & $0.755\pm0.016$        \newrow
$\expval*{\hat{\mathcal{M}}_1}$ & $-1$    &  $-0.11\pm 0.04$    \newrow
$\expval*{\hat{\mathcal{M}}_2}$ & $+1$    &  $0.11\pm 0.04$     \newrow
$\expval*{\hat{\mathcal{M}}_3}$ & $-1$    &  $-0.05\pm 0.04$     \newrow
$\expval*{\hat{\chi}}=\frac{1}{3}\sum\limits_{k=1}^3(-1)^k\hat{\mathcal{M}}_k$& $1$    &  $0.09\pm0.02$\newrow
$\fidelity_{GHZ}^{N=3}=\frac{\expval*{\hat{P}_z}+\expval*{\hat{\chi}}}{2}$             & $100\%$    &  $(42.3\pm1.4)\%$         \newrow
\end{tabular}
\caption{Fidelity estimates using raw measurements of a three-qubit GHZ state. }
\label{tab:3qubit}
\end{table}

\clearpage
\section{Effect of Spectral Diffusion and Blinking}
Spectral diffusion and blinking are observed in the studied quantum dot, yet we exclude these effects from our analysis for the following reasons:
\begin{enumerate}
\item Spectral diffusion mainly affects the fidelity of spin initialization and readout by randomly detuning the cycling transition from the pumping laser. The optical $\pi$-pulse is largely unaffected as the diffusion magnitude $\sigma/2\pi\approx \SI{345}{MHz}$ \cite{Appel2021} is small compared  to the $\pi$-pulse bandwidth $\approx 0.44/T_{opt}\approx\SI{13}{GHz}$ where 0.44 is the Gaussian time-bandwidth product. 
\item Due to the long ms-timescale of spectral diffusion \cite{Kuhlmann2013}, subsequently emitted photons will remain indistinguishable as explicitly demonstrated in recent experiments \cite{Uppu2020}. Reducing spectral diffusion is however important for fusing cluster states generated by separate QDs.
\item Blinking reduces the protocol success probability but does not appreciably reduce the entanglement fidelity due to our post-selection criteria. If (as we expect) the blinking is due to a probabilistic hole initialization at the cycle onset, it only results in a linear decrease in success probability.
\item Finally, recent experiments continue to show improvements to blinking and spectral diffusion. Here we highlight the observation of a virtually blinking-free QD embedded in a PCW \cite{Uppu2020} and the interference of identical photons from independent QDs with <10\% inhomogeneous broadening \cite{Zhai2021}.
\end{enumerate}

\newpage
\nocite{apsrev42Control}
\bibliography{EntanglementSI_arxiv_final.bbl}